\documentclass[10pt,conference]{IEEEtran}
\usepackage{cite}
\usepackage{amsmath,amssymb,amsfonts}
\usepackage{algorithmic}
\usepackage{graphicx}
\usepackage{textcomp}
\usepackage{xcolor}
\usepackage[hyphens]{url}
\usepackage{fancyhdr}
\usepackage{hyperref}

\newcommand{\hpcayear}{2026}

\newcommand{\hpcasubmissionnumber}{744}
\title{AutoGNN: End-to-End Hardware-Driven Graph Preprocessing for Enhanced GNN Performance}

\def\hpcacameraready{} 

\newcommand\hpcaauthors{
  Seungkwan Kang$^1$, Seungjun Lee$^1$, Donghyun Gouk$^2$, Miryeong Kwon$^2$, Hyunkyu Choi$^2$, Junhyeok Jang$^2$, \\
  Sangwon Lee$^2$, Huiwon Choi$^1$, Jie Zhang$^3$, Wonil Choi$^4$, Mahmut Taylan Kandemir$^5$, Myoungsoo Jung$^{1,2}$
}
\newcommand\hpcaaffiliation{\textit{Computer Architecture and Memory Systems Laboratory}, KAIST$^1$, Panmnesia, Inc.$^2$,\\
Peking University$^3$, Hanyang University$^4$, Pennsylvania State University$^5$}
\newcommand\hpcaemail{}

\usepackage{subcaption}
\usepackage{array}
\usepackage{tabularx}
\usepackage{booktabs}
\usepackage{multirow}
\usepackage{graphicx}
\usepackage{dblfloatfix} 
\usepackage{setspace}
\usepackage{xcolor}
\usepackage{colortbl}
\usepackage{epstopdf}

\usepackage[ruled,linesnumbered,noend]{algorithm2e}
\DontPrintSemicolon
\SetAlCapFnt{\small}
\SetKwProg{Fn}{\textbf{fn}}{}{end}

\SetCommentSty{mycommentfont}
\makeatletter
  \newcommand{\removelatexerror}{\let\@latex@error\@gobble}
\makeatother

\author{
  \ifdefined\hpcacameraready
    \IEEEauthorblockN{\hpcaauthors{}}
      \IEEEauthorblockA{
        \hpcaaffiliation{} \\
        \hpcaemail{}
      }
  \else
    \IEEEauthorblockN{\normalsize{HPCA \hpcayear{} Submission
      \textbf{\#\hpcasubmissionnumber{}}} \\
      \IEEEauthorblockA{
        Confidential Draft \\
        Do NOT Distribute!!
      }
    }
  \fi
}

\fancypagestyle{camerareadyfirstpage}{%
  \fancyhead{}
  
  \fancyhead[C]{
    \ifdefined\aeopen
    \parbox[][12mm][t]{13.5cm}{\hpcayear{} IEEE International Symposium on High-Performance Computer Architecture (HPCA)}
    \else
      \ifdefined\aereviewed
      \parbox[][12mm][t]{13.5cm}{\hpcayear{} IEEE International Symposium on High-Performance Computer Architecture (HPCA)}
      \else
      \ifdefined\aereproduced
      \parbox[][12mm][t]{13.5cm}{\hpcayear{} IEEE International Symposium on High-Performance Computer Architecture (HPCA)}
      \else
      \parbox[][0mm][t]{13.5cm}{\hpcayear{} IEEE International Symposium on High-Performance Computer Architecture (HPCA)}
    \fi
    \fi
    \fi
    \ifdefined\aeopen
      \includegraphics[width=12mm,height=12mm]{ae-badges/open-research-objects.pdf}
    \fi
    \ifdefined\aereviewed
      \includegraphics[width=12mm,height=12mm]{ae-badges/research-objects-reviewed.pdf}
    \fi
    \ifdefined\aereproduced
      \includegraphics[width=12mm,height=12mm]{ae-badges/results-reproduced.pdf}
    \fi
  }
  \fancyfoot[C]{}
}
\begin{document}
\maketitle

\newcommand{\hpcaheight}{0mm}
\ifdefined\eaopen
\renewcommand{\hpcaheight}{12mm}
\fi

\setstretch{0.935}

\begin{abstract}
  Graph neural network (GNN) inference faces significant bottlenecks in preprocessing, which often dominate overall inference latency. We introduce AutoGNN, an FPGA-based accelerator designed to address these challenges by leveraging FPGA's reconfigurability and specialized components. AutoGNN adapts to diverse graph inputs, efficiently performing computationally intensive tasks such as graph conversion and sampling. By utilizing components like adder trees, AutoGNN executes reduction operations in constant time, overcoming the limitations of serialization and synchronization on GPUs.

AutoGNN integrates unified processing elements (UPEs) and single-cycle reducers (SCRs) to streamline GNN preprocessing. UPEs enable scalable parallel processing for edge sorting and unique vertex selection, while SCRs efficiently handle sequential tasks such as pointer array construction and subgraph reindexing. A user-level software framework dynamically profiles graph inputs, determines optimal configurations, and reprograms AutoGNN to handle varying workloads. Implemented on a 7$n$m enterprise FPGA, AutoGNN achieves up to 9.0$\times$ and 2.1$\times$ speedup compared to conventional and GPU-accelerated preprocessing systems, respectively, enabling high-performance GNN preprocessing across diverse datasets.

\end{abstract}

\section{Introduction}\label{sec:introduction}
Graph Neural Networks (GNNs) have garnered substantial attention from both industry and academia due to their impressive accuracy across diverse applications \cite{zhou2020graph,yuan2020xgnn,li2022hyperscale}. GNNs achieve superior accuracy over other deep learning methods by effectively learning the features represented by graph vertices and edges, along with their interrelationships \cite{kipf2016semi,velivckovic2017graph}. This capability has made GNNs promising solutions for recommendation systems \cite{wu2020graph,lei2020interactive,he2020lightgcn}, social networks \cite{fan2019graph,fan2020graph,han2020graph}, and knowledge graphs \cite{nathani2019learning,zhang2020relational,yu2021knowledge}.
GNNs integrate graph processing and deep learning operations into a unified learning process, enabling them to handle non-Euclidean data structures \cite{ji2020graph,wu2020comprehensive,asif2021graph}. However, this combination results in inefficiencies when using conventional processing units like CPUs or GPUs \cite{wang2019deep,fey2019fast}. To mitigate this, numerous hardware acceleration techniques and system framework designs have been introduced. Hardware approaches typically address these inefficiencies by i) designing domain-specific accelerators that combine vector and systolic arrays \cite{yan2020hygcn} or ii) utilizing heterogeneous distributed systems incorporating CPUs and GPUs \cite{cai2021dgcl,zhang2020agl,zheng2022distributed}. Meanwhile, system-oriented studies offer frameworks to support parallel execution of GNN processing and deep learning tasks \cite{ma2019neugraph,chen2020fusegnn,wang2021gnnadvisor} via user-friendly programming models \cite{wang2019deep,fey2019fast}.

\enlargethispage{10pt}

Despite these substantial advancements, GNNs remain challenging to deploy in real-world systems due to low service performance. Specifically, we find that preprocessing overhead for large graph datasets accounts for 90.8\% of the total GNN service time from an end-to-end perspective (cf. Section \ref{sec:motivation}).
GNN preprocessing primarily involves \emph{graph conversion} and \emph{graph sampling}, which pertain to transforming the format of the graph dataset and generating a subset of the input graph, respectively. GNN preprocessing is critical in GNN services; without it, \emph{node explosion} can increase the computing demands exponentially \cite{hamilton2017inductive,chen2018fastgcn,ying2018graph,zeng2019graphsaint}, preventing the system from meeting service-level latency agreements (SLAs) in real-world inference. Although some studies propose efficient graph formats or sampling algorithms, the preprocessing overhead itself remains to be addressed in the GNN research domain.


We propose \emph{AutoGNN}, a fully automated preprocessing hardware designed toward enhancing GNN inference performance. AutoGNN executes the entire preprocessing workflow, from start to finish, directly in hardware, producing a subgraph optimized for use by GPUs or other GNN accelerators.
In this work, we design AutoGNN within an FPGA environment to leverage two essential aspects of GNN preprocessing. First, the computational load in GNN preprocessing varies significantly with the characteristics of the input graph. For instance, graphs with numerous relationships require substantial computation for graph conversion, while preprocessing for smaller-scale data is often dominated by graph sampling tasks. Second, GNN preprocessing involves extensive result reduction processes with complex synchronization operations (e.g., locks and atomic transactions). Unfortunately, this often forces GNN preprocessing into a serialized execution, making it challenging to process efficiently on general-purpose GPUs. FPGAs are well-suited to handle these computational dynamics by adapting to varied graph inputs and efficiently executing reduction operations. Its specialized components, such as adder tree logic, enable result reduction in $\mathcal{O}(1)$ time, significantly improving preprocessing performance.

\enlargethispage{10pt}

AutoGNN specifically comprises \emph{Unified Processing Elements} (UPEs) and \emph{Single-Cycle Reducers} (SCRs).
The UPE is designed to perform two graph-specific operations -- edge sorting and unique vertex selection -- within a single hardware logic, each operation corresponding to graph conversion and sampling, respectively.
Since each UPE can handle both edges and vertices through shared hardware logic, AutoGNN can scale with multiple UPEs based on input graph demands.
This design maximizes FPGA resource utilization, achieving a high level of parallelization and bandwidth efficiency with limited resources. In contrast, SCRs handle the non-parallelizable tasks in both graph conversion and sampling, further optimizing GNN preprocessing efficiency.

We also provide user-level software that identifies the characteristics of GNN preprocessing by capturing the dynamics of the input graph. Based on this analysis and a cost function, the software determines the optimal hardware configuration and reprograms the underlying AutoGNN accordingly. We implemented all hardware modules of AutoGNN on a 7$n$m enterprise FPGA evaluation board and evaluated end-to-end GNN processing to explore the design space with various system configurations. Evaluation results show performance improvements of 9.0$\times$ and 2.1$\times$ compared to conventional and GPU-accelerated preprocessing systems, respectively.

The main contributions of this paper are as follows:

\begin{figure}
  \centering
  \vspace{0pt}
  \begin{minipage}[t]{0.55\linewidth}
    \includegraphics[width=\linewidth]{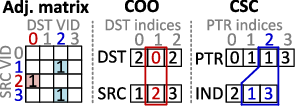}
    \vspace{-10pt}
    \renewcommand*{\arraystretch}{0.3}
    \caption{Graph representation.}\label{fig:back_graph_repr}
  \end{minipage}%
  \begin{minipage}[t]{0.4\linewidth}
    \includegraphics[width=\linewidth]{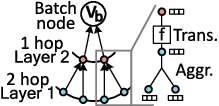}
    \vspace{-10pt}
    \renewcommand*{\arraystretch}{0.3}
    \caption{GNN inference.}\label{fig:back_graph_infer}
  \end{minipage}
  \vspace{-15pt}
\end{figure}

\noindent $\bullet$ \emph{Comprehensive analysis and characterization of GNN preprocessing.} Although GNN preprocessing is the most time-consuming component in GNN services, there is a lack of research on the specific tasks within this process and their performance bottlenecks. In this study, we break down GNN preprocessing into distinct tasks, examining them both qualitatively and quantitatively. Our in-depth analysis classifies these tasks into parallelizable and non-parallelizable categories, enabling targeted acceleration strategies within a single preprocessing hardware architecture.

\noindent $\bullet$ \emph{Unified hardware design for parallelizable tasks.}
We reveal that edge sorting and unique vertex selection in GNN preprocessing can be efficiently addressed using prefix-sum \cite{blelloch1990prefix,harris2007parallel} and routing \cite{Chang1997ArbitrarySB,andrey20sort} algorithms. In addition, we observe that both algorithms can be combined to support set-partition operations \cite{Aumuller2018SimpleAF,Kiwiel2003PartitioningSF}, which involve extracting data to form a set that meets given conditions. Leveraging these characteristics, we design UPEs, each capable of executing parallelizable tasks within a single hardware architecture. This unified approach maximizes hardware resource utilization, thereby enhancing computing bandwidth.


\noindent $\bullet$ \emph{Acceleration of non-parallelizable tasks.}
Graph conversion and sampling in GNN preprocessing often require counting specific vertices or edges, tasks that are typically handled serially or with mutual exclusion. To overcome this limitation, we design the SCR with multiple comparator logic, enabling parallel comparisons across numerous inputs. The SCR then aggregates these results using an adder tree or a filter tree, performing vertex or edge counting in a single cycle. By deploying SCRs alongside UPEs in configurations tailored to the input graph's requirements, AutoGNN autonomously executes the entire preprocessing workflow with high efficiency.

\enlargethispage{10pt}

\noindent $\bullet$ \emph{Reconfigurable design for dynamic graphs.}
While UPEs and SCRs are optimized for efficient hardware utilization, the resource demands of GNN preprocessing can vary significantly with graph characteristics. A fixed configuration may not always provide optimal performance. To address this, we implement a cost model that evaluates hardware configurations and graph features to identify the optimal balance of UPEs and SCRs. AutoGNN dynamically reconfigures the FPGA at runtime only when the model determines it is necessary. To minimize reconfiguration overhead, we analyze each parameter's impact on reprogramming area and latency, allowing it to update only the essential hardware modules selectively.

\section{Background}\label{sec:background}
\subsection{Elements of GNN Processing and Service}
\label{sec:back-gnn}

\noindent \textbf{Graph representation.}
The basic graph representation is the adjacency matrix, where row and column indices represent source and destination \emph{vertex identifications} (VIDs), respectively. Each element indicates whether the two corresponding vertices are connected (1) or not (0), resulting in a matrix with many zero entries, which is inefficient \cite{cornell20graph,Huang2020GESpMMGS}. To enhance the practicality of adjacency matrices in graph processing, various optimized formats have been developed \cite{Hoefler2021SparsityID,wang2019deep}. This section briefly introduces two common representations: \emph{coordinate format} (COO) and \emph{compressed sparse column} (CSC), as shown in Figure \ref{fig:back_graph_repr}. The COO format, often called an edge array, stores each edge as a pair of source VID and destination VID, all in an unsorted manner \cite{Abi-Karam2022GenGNNAG}.
In contrast, CSC is a vertex-centric structure enabling efficient access to edges associated with a particular node \cite{Zhang2021GammaLG,Saad2003IterativeMF}.
For example, with a given destination VID, CSC can quickly identify all connected source nodes, retrieving all edges associated with that destination. As shown in Figure \ref{fig:back_graph_repr}, CSC consists of two arrays: pointers and indices. The \emph{pointer array}'s indices correspond to the columns of an adjacency matrix (or destination VIDs), with each value indicating the start offset in the index array. The \emph{index array}, in turn, contains row indices (or source VIDs). To retrieve all source VIDs connected to a destination VID, one can simply access the range of elements in the index array, starting from the offset of the pointer array up to the next pointer. While COO is suitable for frequently updated graphs, CSC provides efficient graph traversal.

\noindent \textbf{Neural network processing with graphs.}
In graph theory, transitioning from one node to another via an edge is referred to as a \emph{hop} \cite{Cohen2002ReachabilityAD,zhou2020graph}. A GNN is composed of multiple layers, each representing the one-hop neighborhood of a vertex. These neighborhoods encompass the nodes directly connected to a given \emph{batch node}, which acts as the starting point for inference.
To infer features of the batch node, the GNN aggregates the embeddings of neighboring nodes in each layer during an \emph{aggregation} step. The combined embedding is then passed through a conventional deep neural network (DNN), which transforms it to a representation for the next hop (called a \emph{transformation} step). This aggregation-transformation cycle is repeated for all layers, enabling GNNs to learn not only the node and edge information but also the relationships between them \cite{yan2020hygcn,You2021GCoDGC}. This allows GNNs to achieve higher accuracy than traditional DNNs \cite{velivckovic2017graph,hamilton2017inductive,kipf2016semi}.

\enlargethispage{10pt}

Figure \ref{fig:back_graph_infer} shows an example graph spanning two GNN layers and the corresponding inference process, respectively.
The inference begins with the layer farthest from the batch node ($V_b$), represented as the 2-hop layer (layer 1). For each layer, the embeddings of neighboring nodes are aggregated into a single embedding.
This aggregated embedding is then transformed into a new representation for the next layer using a function $f$. For the last layer (layer 2), the transformation is typically performed through a multi-layer perceptron (MLP), generating a new embedding that captures all relationships within the subgraph around the batch node. This final embedding serves as the inference result.

\noindent \textbf{GNN preprocessing for inference.}
Most graph traversals in GNN services depend on identifying destination nodes from source nodes, making CSC format preferable for efficient traversal.
However, raw or application-specific graphs are often stored in COO format for storage efficiency and graph update flexibility \cite{Abi-Karam2022GenGNNAG,Chen2023DGNNBoosterAG}. As a result, converting graphs from COO to CSC format is an essential preprocessing step before inference \cite{Mostafa2023FastSampleAD}. This \emph{graph conversion} is one of the most time-consuming tasks in GNN preprocessing.
Another crucial task in GNN preprocessing is \emph{graph sampling} \cite{Kaler2021AcceleratingTA}.
Growing graph datasets require substantial memory and increase inference latency.
Specifically, as the number of layers and node degree increase, the number of nodes that GNN must explore grows exponentially, a phenomenon known as \emph{node explosion} \cite{Tripathy2020ReducingCI,Duan2022ACS}.
The problem amplifies if a node with a high degree (number of neighbor nodes) is accessed during traversal. For example, the Movie dataset may require traversing 99\% of the total graph for a two-layer GNN, depending on the batch node.
Node explosion leads to significant and unpredictable latency increases in GNN inference, limiting its application to autonomous driving \cite{shi2020point}, high-energy physics \cite{shlomi2020graph}, and recommendation systems \cite{chantat2018pixie, dandan2022platogl}.
To mitigate this, studies proposed to sample a subset of the original graph before inference \cite{jie2025helios, shuangchen2022hyperscale, tim2022accelerating}.
For instance, \cite{hamilton2017inductive} samples a fixed number of nodes per hop, which greatly reduces the computation and memory requirements while preserving reasonable accuracy.

\enlargethispage{10pt}

\begin{figure}
  \centering
  \vspace{-10pt}
  \includegraphics[width=\linewidth]{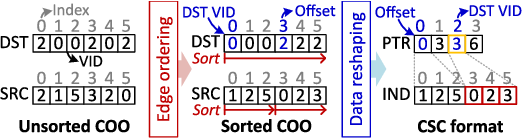}
  \vspace{-2pt}
  \caption{Graph conversion.}\label{fig:back_graph_conv}
  \vspace{-32pt}
  \begin{subfigure}{1\linewidth}
    \renewcommand*{\arraystretch}{0.3}
    \begin{tabularx}{\textwidth}{
      p{\dimexpr.5\linewidth-1.33\tabcolsep}
      p{\dimexpr.5\linewidth-1.33\tabcolsep}
      }
      \caption{Edge ordering.}\label{fig:back_graph_conv_1} &
      \caption{Data reshaping.}\label{fig:back_graph_conv_2}
    \end{tabularx}
  \end{subfigure}
  \vspace{-10pt}
\end{figure}

\subsection{Decomposition of GNN Preprocessing}\label{subsec:decompose_prep}

Graph conversion and sampling introduce additional operations into GNN services that were not previously required, placing them on the critical path of GNN processing \cite{Liu2021GNNSamplerBT}. Graph conversion primarily involves two tasks: \emph{edge ordering} and \emph{data reshaping}, while graph sampling consists of \emph{unique random selection} (uni-random selection) and \emph{subgraph reindexing}. Edge ordering and data reshaping process a large number of graph components, necessitating extensive parallelization to improve efficiency. In contrast, uni-random selection and subgraph reindexing work on a smaller subset of graph elements but require frequent updates and synchronization, making these more complex and computationally demanding.

\noindent \textbf{Edge ordering.}
Edge ordering facilitates efficient access to nodes based on destination VIDs. It begins by sorting edges primarily by their destination VIDs and then secondarily by their source VIDs, maintaining the order of destinations. This approach ensures that edges sharing the same destination node are grouped sequentially in the result.
As shown in Figure \ref{fig:back_graph_conv}, this sorted edge array serves as a foundational structure for the CSC format, enhancing the efficiency and performance of graph sampling and aggregation operations in GNN inference.

\noindent \textbf{Data reshaping.}
While edge ordering produces a sorted COO array, locating all source nodes for a given destination VID still requires a binary search. To overcome this, data reshaping repurposes the sorted COO array into an index array, creating range information for each group of edges that share the same destination VID. The pointer array is constructed by scanning the index array sequentially and setting a start offset whenever a new destination VID is encountered. This process is computationally intensive and offers limited opportunities for optimization. We will examine its performance challenges in more detail in the following section.

\begin{figure}
  \centering
  \vspace{-10pt}
  \includegraphics[width=\linewidth]{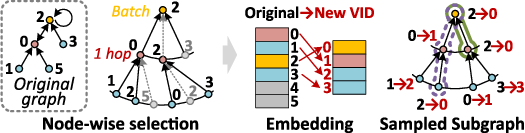}
  \vspace{0pt}
  \caption{Graph sampling.}\label{fig:back_graph_samp}
  \vspace{-32pt}
  \begin{subfigure}{1\linewidth}
    \renewcommand*{\arraystretch}{0.3}
    \begin{tabularx}{\textwidth}{
      p{\dimexpr.45\linewidth-1.33\tabcolsep}
      p{\dimexpr.55\linewidth-1.33\tabcolsep}
      }
      \caption{Unique random selection.}\label{fig:back_graph_samp_1} &
      \caption{Subgraph reindexing.}\label{fig:back_graph_samp_2}
    \end{tabularx}
  \end{subfigure}
  \vspace{-10pt}
\end{figure}

\noindent \textbf{Unique random selection.} This is a critical step in sampling a subgraph and can be performed in two ways: node-wise or layer-wise selection \cite{Vatter2023TheEO,Li2024CeleritasOB}. In node-wise selection (Figure \ref{fig:back_graph_samp_1}), a batch node selects $k$ neighbors at the first hop, which are iteratively sampled for additional $k$ neighbors across multiple hops to expand the neighborhood. In contrast, layer-wise selection samples $k$ neighbors at each layer without requiring interconnections, completing the process in fewer steps. While layer-wise selection is faster, node-wise selection is preferred for its higher accuracy.
Ensuring uniqueness and randomness during the vertex selection process is crucial. Uniqueness prevents redundant selections, preserving the total number of sampled vertices, while randomness improves inference accuracy \cite{hamilton2017inductive,Liu2021SamplingMF}. This process often involves checking a synchronized dictionary (i.e., map) to track selected nodes.

\enlargethispage{10pt}

\noindent \textbf{Subgraph reindexing.} After sampling the original graph, the embedding features must be restructured to align with the sampled nodes. As shown in Figure \ref{fig:back_graph_samp_2}, this involves generating a new embedding table by extracting the embeddings of the sampled vertices from the original embedding table, which is ordered by VIDs. However, since the indices in the sampled embedding table differ from the original VIDs, renumbering these VIDs is required to maintain consistency. Subgraph reindexing addresses this by mapping each original graph VID to a new VID in the sampled subgraph.
Although this procedure has lower latency than other GNN preprocessing tasks, the requirement to manage mapping information in a mutually exclusive manner introduces additional delays.
Note that while each hop of the uni-random selection ensures newly chosen vertices are distinct, loops in the parent-child relationships may lead to repeated vertices in the final result. Specifically, a vertex can reconnect to itself (highlighted by a solid line) or be revisited after multiple hops (highlighted by a dashed line). Consequently, subgraph reindexing outputs are initially collected in COO format, then undergo edge ordering and data reshaping to produce the final CSC representation of the sampled subgraph.

\section{Challenge and Motivation}\label{sec:motivation}

\subsection{Analysis of GPU-Augmented Preprocessing}
\label{sec:motiv-gpu-preprocessing}

To better understand the overhead imposed by GNN preprocessing, we evaluate 11 real-world graph datasets selected from open-source benchmarks~\cite{hu2020ogb,wang2019deep,fey2019fast}. Table~{\ref{tbl:datasets_summary}} summarizes the key characteristics of graph datasets. The preprocessing and inference tasks are executed on an RTX 3090 GPU~{\cite{rtx3090}} using DGL~{\cite{wang2019deep}}, a state-of-the-art GNN framework.

\begin{figure}
  \centering
  \vspace{-10pt}
  \begin{minipage}[t]{0.49\linewidth}
    \vspace{0pt}
    \includegraphics[width=\linewidth]{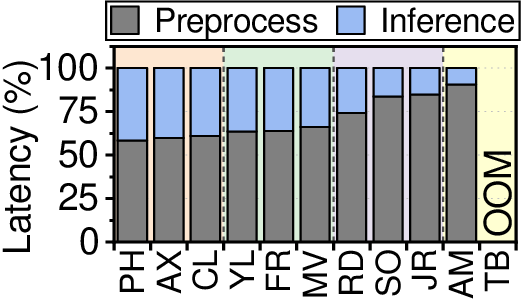}
    \vspace{-13pt}
    \caption{Analysis of GNN\\ preprocessing overhead.}\label{fig:motiv_large_preprocess}
  \end{minipage}%
  \begin{minipage}[t]{0.49\linewidth}
    \vspace{0pt}
    \includegraphics[width=\linewidth]{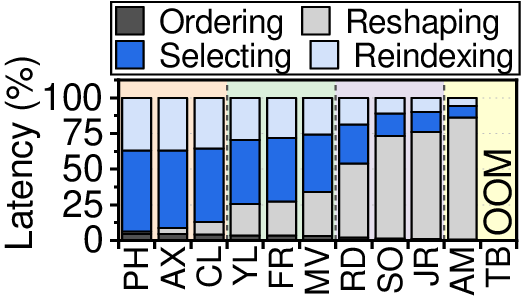}
    \vspace{-13pt}
    \caption{Breakdown analysis of GNN preprocessing.}\label{fig:motiv_gpu_breakdown}
  \end{minipage}%
  \vspace{-15pt}
\end{figure}

\noindent \textbf{Analyzing overhead significance.}
Figure \ref{fig:motiv_large_preprocess} shows the proportion of GNN preprocessing within the total latency of GNN services. For clarity, workloads from each domain are arranged from left to right in ascending order of edge count. Despite the use of GPU acceleration in DGL to optimize preprocessing, it still accounts for an average of 70\% of the total inference time. The figure also reveals that the overhead of GNN preprocessing becomes increasingly significant as the graph size grows. This is because preprocessing load scales proportionally with graph size, whereas inference latency (\texttt{Inference}) remains relatively stable due to graph sampling.

\noindent \textbf{Decomposing the preprocessing latency.} Figure \ref{fig:motiv_gpu_breakdown} breaks down the latency of GNN preprocessing into its four main tasks: edge ordering (\texttt{Ordering}), data reshaping (\texttt{Reshaping}), unique vertex selection (\texttt{Selecting}), and subgraph reindexing (\texttt{Reindexing}). Two key observations emerge from the analysis. First, no single task consistently dominates preprocessing time across all network domains. Second, latency trends vary with the size of the input graphs.

Specifically, for smaller graphs with fewer than 500K edges, graph sampling (\texttt{Selecting} + \texttt{Reindexing}) contributes the most to preprocessing latency, with \texttt{Selecting} and \texttt{Reindexing} accounting for 33.8\% and 22.1\% of the latency, respectively. However, their contributions decrease by 7.5\% and 5.2\% as the graph size increases. This reduction occurs because \texttt{Selecting} applies to a limited number of hops, regardless of graph size.
In contrast, for larger graphs with edge counts ranging from tens of millions to billions, graph conversion (\texttt{Ordering} + \texttt{Reshaping}) becomes the primary bottleneck.
In these cases, \texttt{Ordering} and \texttt{Reshaping} account for 1.8\% and 86.1\% of the latency, respectively.
This shift is due to the need for graph conversion to process the entire graph, both the original graph for sampling and the sampled subgraph for inference.

\begin{figure}
  \vspace{-10pt}
  \includegraphics[width=\linewidth]{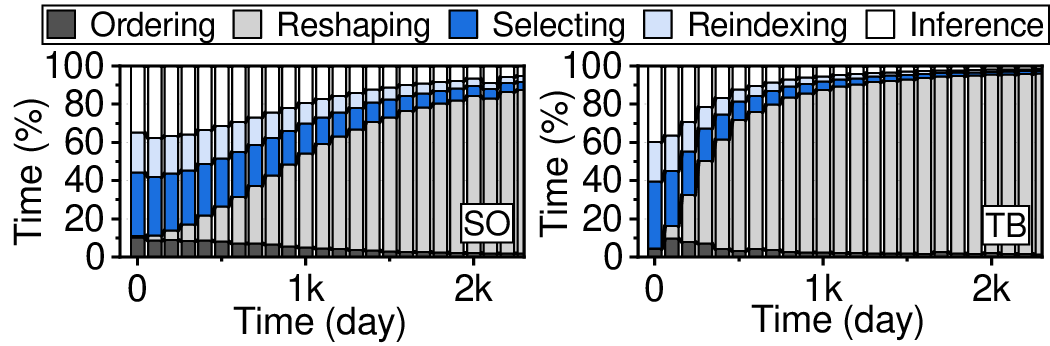}
  \vspace{-6pt}
  \caption{Latency breakdown of dynamic graphs.}\label{fig:motiv_dynamic}
  \vspace{-30pt}
  \begin{subfigure}{1\linewidth}
    \renewcommand*{\arraystretch}{0.3}
    \begin{tabularx}{\textwidth}{
      p{\dimexpr.47\linewidth-1.33\tabcolsep}
      p{\dimexpr.52\linewidth-1.33\tabcolsep}
      }
      \caption{Social network (SO).}\label{fig:motiv_dynamic_degree} &
      \caption{E-commerce (TB).}\label{fig:motiv_dynamic_breakdown}
    \end{tabularx}
  \end{subfigure}
  \vspace{-10pt}
\end{figure}

\noindent \textbf{Considering graph dynamics.} Even within a single workload, the latency trends for the four primary preprocessing tasks can change significantly over time. Figure \ref{fig:motiv_dynamic} shows GNN service latency for two large datasets, SO and TB, which represent social and e-commerce network domains, respectively. In these domains, user-to-user connections and behaviors are frequently updated, causing the graphs to evolve continuously. For example, the number of edges in each dataset increases by 0.52\% and 0.95\% per day, respectively, highlighting the dynamic nature of these graphs.
As shown in Figure \ref{fig:motiv_dynamic}, the latency initially is dominated by \texttt{Selecting} within the GNN inference services. Over time, however, this trend shifts. After 400 days (SO) and 20 days (TB), respectively, \texttt{Reshaping} becomes increasingly significant in terms of latency, eventually surpassing \texttt{Selecting} as the dominant task. This shift occurs because \texttt{Reshaping} must handle the continuously growing number of edges and degrees, while the latency of \texttt{Selecting} remains bounded by the fixed $k$ value. This pattern is consistent across both datasets and underscores the need for adaptable hardware configurations, as no single fixed setup can accelerate all tasks under such dynamic conditions.

\begin{figure*}
  \vspace{-10pt}
  \begin{minipage}{0.3\linewidth}
    \includegraphics[width=\linewidth]{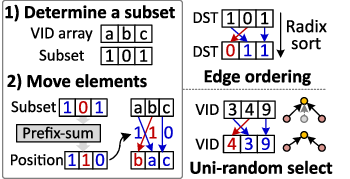}
    \vspace{-8pt}
    \caption{Set-partitioning.}\label{fig:set_partitioning}
  \end{minipage}
  \begin{minipage}{0.32\linewidth}
    \includegraphics[width=\linewidth]{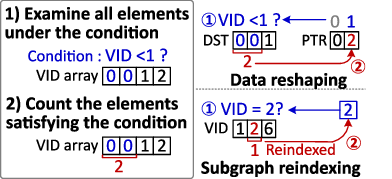}
    \vspace{-8pt}
    \caption{Set-counting.}\label{fig:set_counting}
  \end{minipage}
  \begin{minipage}{0.34\linewidth}
    \includegraphics[width=\linewidth]{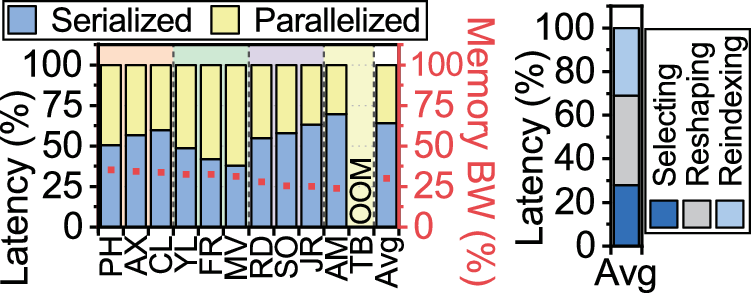}
    \vspace{-5pt}
    \caption{Serialized computation analysis.}\label{fig:motiv_atomic_time_all}
    \vspace{-30pt}
    \begin{subfigure}{1\linewidth}
      \renewcommand*{\arraystretch}{0.3}
      \begin{tabularx}{\textwidth}{
        p{\dimexpr.6\linewidth-1.33\tabcolsep}
        p{\dimexpr.4\linewidth-1.33\tabcolsep}
        }
        \caption{Execution breakdown.}\label{fig:motiv_atomic_time_all_datasets} &
        \caption{Serial tasks.}\label{fig:motiv_atomic_time_all_breakdown}
      \end{tabularx}
    \end{subfigure}
  \end{minipage}
  \vspace{-10pt}
\end{figure*}

\subsection{Removing Preprocessing from Critical Path}
\label{sec:motiv-critical-remove}
To eliminate preprocessing overhead from the critical path of GNN services, AutoGNN accelerates and fully automates all time-consuming tasks in hardware, streamlining the entire preprocessing workflow. This is achieved by \emph{fundamentally redesigning the underlying algorithms to minimize mutual exclusion (atomic) operations in GNN preprocessing, while maximizing parallelization} and acceleration wherever possible.
AutoGNN’s efficiency stems from its core design, which incorporates \emph{unified processing elements} (UPEs) and \emph{single-cycle reducers} (SCRs). These reconfigurable components allow for varying logic counts and hardware sizes to accommodate diverse graph preprocessing needs. UPEs efficiently handle edge ordering and unique random selection, while SCRs significantly reduce the latency of data reshaping and subgraph reindexing. Together, these components ensure high performance across a wide range of GNN network domains.


\noindent\textbf{Accelerating \texttt{Ordering} and \texttt{Selecting}.}
UPE is designed with the insight that both edge ordering and uni-random selection can be implemented using a common operation: extracting elements that satisfy specific conditions and organizing them into a new set~\mbox{\cite{Aumuller2018SimpleAF,Kiwiel2003PartitioningSF}}, called \emph{set-partitioning} in this work.
For edge ordering, UPE leverages a characteristic of typical graph datasets: although there are many vertices, their VIDs are integers drawn from a small, contiguous range.
We therefore target radix sort, whose digit-wise passes are precisely set-partitioning; this makes it an ideal fit for UPE, and accelerating set-partitioning with UPE directly speeds up radix sort.
For uni-random selection, UPE first uses set-partitioning to split the vertices into two buckets, one containing the sampled vertices and the other containing the unsampled vertices. It then repeatedly draws a random vertex from the unsampled bucket only, guaranteeing uniqueness without a full-space scan, thereby improving parallelism and performance.

A key challenge in set-partitioning is determining the new position of each extracted element in the output set. UPE addresses this by integrating two logic blocks: one calculates the cumulative sums for the elements in an array, indicating the distance each element should shift to the left, and the other repositions the elements based on these offsets, all in a single cycle. This enables UPE to partition sets by radix or state (e.g., sampled or unsampled) in just a few cycles. Further details on UPE are discussed in Section \ref{subsec:impl-upe}.

\noindent \textbf{Reducing \texttt{Reshaping} and \texttt{Reindexing}.} Data reshaping and subgraph reindexing can also be both reduced to a single core operation: counting elements in a set that satisfy a given condition, a process referred to as \emph{set-counting} in this work. The primary challenge in implementing this algorithm lies in its inherently mutually exclusive nature. Updating a counter variable typically requires atomic operations, which severely limit parallelism in these preprocessing tasks. In addition, the processing bandwidth for data reshaping and subgraph reindexing is constrained by the size of the input array and the number of threads available for simultaneous operation.

The SCR overcomes these challenges by utilizing thousands of comparators and an adder/filter tree to aggregate the comparator outputs in a single cycle. This architecture enables efficient computation of the range information required for constructing the pointer array in CSC format, thereby significantly accelerating \texttt{Reshaping}. Similarly, SCR identifies specific VIDs in the input, streamlining the creation of an index map and thereby reducing the complexity of \texttt{Reindexing}. Further details on SCR's design and implementation are discussed in Section \ref{sec:impl-scr}.

\enlargethispage{10pt}

\section{Hardware-Driven GNN Preprocessing}\label{sec:overview}
\subsection{Redesigning GNN Preprocessing}\label{sec:overview-hardware}

\noindent \textbf{Set-partitioning.}
This is an algorithm that divides a given array of VIDs into two disjoint subsets by evaluating each element, determining its subset, and moving the element accordingly. Instead of scanning all elements, set-partitioning can be efficiently implemented by relocating elements based on prefix-sum results. The prefix-sum operation generates a new array by cumulatively counting elements in the input array that satisfy a specific condition, allowing us to evaluate the position of each element in the new set.
Figure~\mbox{\ref{fig:set_partitioning}} shows set-partitioning used in edge ordering and uni-random selection.
Each prefix-sum value is the element's exclusive write index in the output: for
edge ordering, it gives the radix bucket offsets, and for uni-random selection, it gives the compact position in the unsampled array. Using these indices, we scatter elements in one pass, streamlining both tasks.


\noindent \textbf{Set-counting.}
This operation examines all elements in a set against a specified condition and counts the number of elements that satisfy it. As shown in Figure \ref{fig:set_counting}, it is applicable to both data reshaping and subgraph reindexing. In data reshaping, the pointer array is constructed based on the result of edge ordering, which consists of arrays for destination and source VIDs.
Data reshaping can be viewed as set-counting based on the observation that the index of the pointer array corresponds directly to the destination VID, and the value at each index represents the number of elements with destination VIDs smaller than the index.
Thus, set-counting can efficiently populate the pointer array by counting the elements that satisfy the condition for each destination VID.
Note that a common approach is to increment a counter for the processed edges and update the previous edge's destination VID while traversing the sorted edge array. However, this requires sequential execution, as each step depends on the previous one. Viewing data reshaping as set-counting effectively enables concurrent computation of each pointer array entry, thus improving throughput.


In subgraph reindexing, the goal is to map the results of uni-random selection to renumbered IDs if they have not been mapped already. The critical step in this process is checking and updating the mapping. While a hash map is typically used for this process \cite{Carter2019NanosecondIO}, resizing the hash map incurs a time complexity of $O(n)$. Set-counting can replace the hash map in this process. Similar to data reshaping, two arrays are maintained: one for the original VIDs (before reindexing) and another for the updated VIDs (after reindexing). By setting the VID from uni-random selection as the condition for set-counting, it can determine whether the VID has been reindexed without relying on a hash map, streamlining the process.

\enlargethispage{10pt}


\noindent \textbf{Limits for redesigned algorithms.}
Although the redesigned algorithms for set-partitioning and set-counting can be accelerated on conventional GPUs, their parallel architectures cannot fully remove GNN preprocessing from the critical path of GNN services. These limitations arise from the need to synchronize shared resources, such as the counter variables and map structures, across multiple GPU threads, which restricts the simultaneous execution of all threads. As shown in Figure \ref{fig:motiv_atomic_time_all_datasets}, when GNN preprocessing is accelerated on a GPU (RTX 3090) using set-partitioning and set-counting within a CUDA kernel, the execution time is divided into parallelized and serialized portions. As shown in the figure, 64.1\% of the overall execution time remains serialized, on average, resulting in low GPU resource utilization.
Specifically, only 30.3\% of the GPU's memory bandwidth is utilized on average, negatively impacting performance. Further analysis in Figure~\ref{fig:motiv_atomic_time_all_breakdown} shows that uni-random selection, data reshaping, and subgraph reindexing contribute 27.9\%, 41\%, and 31.1\%, respectively, on average, to the non-parallelizable tasks.

\subsection{System Architecture Overview}
To overcome these challenges, AutoGNN optimizes set-partitioning and set-counting by minimizing non-parallelizable tasks while maximizing resource utilization on FPGAs through UPEs and SCRs. While FPGAs are generally slower than ASICs, AutoGNN's hardware design achieves high performance in graph dataset preprocessing by fully exploiting FPGA resources. Its modular architecture, with configurable UPEs and SCRs, allows seamless adaptation to dynamic graphs and varying workloads. In addition, the accompanying software framework enhances flexibility by dynamically adjusting hardware configurations at runtime. By decoupling processing bandwidth from dataset-specific constraints, AutoGNN consistently delivers efficient and robust preprocessing for GNN services across diverse environments.

\begin{figure}
  \centering
  \vspace{-10pt}
  \includegraphics[width=\linewidth]{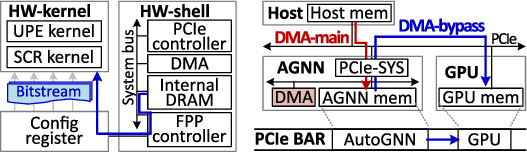}
  \vspace{2pt}
  \caption{High-level view of AutoGNN.}\label{fig:overall}
  \vspace{-32pt}
  \begin{subfigure}{1\linewidth}
    \renewcommand*{\arraystretch}{0.3}
    \begin{tabularx}{\textwidth}{
      p{\dimexpr.5\linewidth-1.33\tabcolsep}
      p{\dimexpr.5\linewidth-1.33\tabcolsep}
      }
      \caption{Overall architecture.}\label{fig:overall_architecture} &
      \caption{AutoGNN interface.}\label{fig:AutoGNN_interface}
    \end{tabularx}
  \end{subfigure}
  \vspace{-10pt}
\end{figure}

Figure \ref{fig:overall_architecture} shows the overall architecture of AutoGNN, which comprises two main FPGA components: the hardware kernel (\emph{HW-kernel}) and the hardware shell (\emph{HW-shell}). The HW-kernel is a reconfigurable area containing the UPE kernel and SCR kernel, while the HW-shell is fixed and unaffected by changes in the input graph dataset. The HW-shell includes peripheral components such as the FPGA programming port (\emph{FPP}) controller, for example, ICAP \cite{Pezzarossa2017ACF,Lai2009ICAPIAR}, and a PCIe controller. These modules are interconnected via a system bus, allowing the host to reconfigure UPEs and SCRs on the HW-kernel by writing pre-prepared bitstreams from AutoGNN's internal DRAM to the FPP controller and associated registers.

\enlargethispage{10pt}

Figure \ref{fig:AutoGNN_interface} depicts how AutoGNN interfaces with the host and GPUs (or other types of accelerators). A PCIe-SYS controller serves as the bridge between AutoGNN's internal system bus and the PCIe interface, exposing two distinct DMA regions: DMA-main and DMA-bypass. DMA-main facilitates data transfer using a PCIe descriptor, which acts as a scatter-gather list to enable AutoGNN to efficiently copy large-scale COO data from the host's system memory. This is particularly useful for COO datasets that are scattered across system memory but contiguous in userland. In contrast, DMA-bypass supports conventional memory reads and writes via PCIe BAR, functioning similarly to memory-mapped I/O (MMIO). The DMA-bypass is accessible to the host’s software framework, enabling direct mapping to the GPU or other accelerators' internal memory. This simplifies the transfer of preprocessed, small-sized results, such as subgraphs, to the host or GPU.

\subsection{Reconfigurable Blocks for Acceleration}\label{subsec:impl-upe}
\begin{figure}
  \vspace{-10pt}
  \centering
  \includegraphics[width=\linewidth]{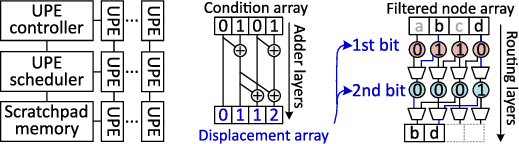}
  \vspace{3pt}
  \caption{UPE kernel design.}\label{fig:over_UPE_kernel}
  \vspace{-32pt}
  \begin{subfigure}{1\linewidth}
    \renewcommand*{\arraystretch}{0.3}
    \begin{tabularx}{\textwidth}{
      p{\dimexpr.3\linewidth-1.33\tabcolsep}
      p{\dimexpr.35\linewidth-1.33\tabcolsep}
      p{\dimexpr.35\linewidth-1.33\tabcolsep}
      }
      \caption{UPE kernel.}\label{fig:over_UPE_kernel_1} &
      \caption{Prefix-sum logic.}\label{fig:over_UPE_kernel_2} &
      \caption{Relocation logic.}\label{fig:over_UPE_kernel_3}
    \end{tabularx}
  \end{subfigure}
  \vspace{-10pt}
\end{figure}

\noindent \textbf{UPE kernel design.}
Figure \ref{fig:over_UPE_kernel_1} shows the internal structure of the UPE kernel, consisting of a UPE controller, multiple UPEs, and a UPE scheduler. These components are interconnected via a crossbar switch and share a scratchpad memory. The UPE controller manages input data and intermediate results, storing and retrieving them from the scratchpad memory. To maximize parallelism, the UPE scheduler oversees UPE execution, using a scoreboard to track the status of each UPE (busy or idle) and assign input data accordingly.

Each UPE integrates prefix-sum and relocation logic, and is reconfigurable in its number and size. It processes two input arrays: one for nodes (VIDs) and another for condition values (booleans). The condition array is fed into the prefix-sum logic, which generates a \emph{displacement array}. This array contains cumulative counts of elements in the node array that satisfy the given condition values. Figure \ref{fig:over_UPE_kernel_2} provides an example of this logic with four input elements.
The prefix-sum logic is implemented as a hierarchical adder network: the first layer computes local sums for the first and second halves of the array, and subsequent layers propagate and combine these sums. The result is a displacement array produced in $O(\log n)$ adder layers. Because the inputs are booleans, each adder only needs a width of $\log n$ bits. We observed that this hardware can process hundreds of elements in a single cycle.

Meanwhile, the node array is processed through AND gates with the condition array, producing a filtered node array where elements that do not meet the conditions are cleared to zero. The displacement array is then fed into the relocation logic, which shifts elements in the filtered node array based on the corresponding values in the displacement array.
As shown in Figure {\ref{fig:over_UPE_kernel_3}}, the relocation logic employs $O(\log n)$ routing layers. Each layer decomposes the movement distances of input elements into powers of two and shifts the elements accordingly, moving them leftward at each stage.
Note that each multiplexer in the relocation logic has a single-bit select input to choose between two sources for the subsequent layer. The input/output width matches the bit width of the array elements being aligned (64 bits in AutoGNN to store two VIDs).
These iterative operations, carried out by the prefix-sum and relocation logic, support various preprocessing tasks such as edge ordering and unique random selection.
The UPE controller orchestrates these steps and combines their outputs to complete the desired preprocessing, depending on the specific task or data size.

\enlargethispage{10pt}

\begin{figure}
  \vspace{0pt}
  \centering
  \includegraphics[width=\linewidth]{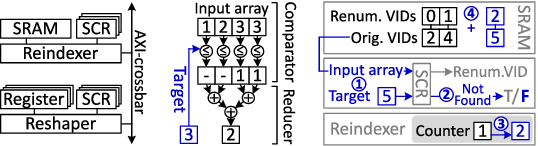}
  \vspace{2pt}
  \caption{SCR kernel design.}\label{fig:over_SCR_kernel}
  \vspace{-33pt}
  \begin{subfigure}{1\linewidth}
    \renewcommand*{\arraystretch}{0.3}
    \begin{tabularx}{\textwidth}{
      p{\dimexpr.26\linewidth-1.33\tabcolsep}
      p{\dimexpr.30\linewidth-1.33\tabcolsep}
      p{\dimexpr.42\linewidth-1.33\tabcolsep}
      }
      \caption{SCR kernel.}\label{fig:over_SCR_kernel_1} &
      \caption{SCR structure.}\label{fig:over_SCR_kernel_2} &
      \caption{Reindexer.}\label{fig:over_SCR_kernel_3}
    \end{tabularx}
  \end{subfigure}
  \vspace{-10pt}
\end{figure}

\noindent \textbf{SCR kernel design.}\label{sec:impl-scr}
Figure \ref{fig:over_SCR_kernel_1} shows the SCR kernel architecture, which comprises two primary controllers: the reshaper (reshaping controller) and the reindexer (reindexing controller). These controllers are connected via a simple bus, such as an AXI crossbar in our implementation, which provides a single port to the HW-shell. This design ensures correct signal timing regardless of the placement of the reshaper and reindexer. The reshaper is equipped with a set of registers to handle the COO format and intermediate results from the underlying SCRs, while the reindexer incorporates an SRAM bank to store the mapping information.
Behind the reshaper and reindexer, a reconfigurable number of SCRs are deployed.

Each SCR is composed of two key components: comparator logic and reducer logic. The comparator evaluates all elements of the input array against a given target, while the reducer aggregates the comparison results into a single output. Figure \ref{fig:over_SCR_kernel_2} depicts the internal structure of an SCR, which can be configured for either reshaping or reindexing tasks. An SCR receives two inputs: an array and a comparison target.
For the reshaper, the comparator subtracts the target from each element of the input array and outputs a positive result if the subtraction is greater than or equal to zero.
Note that the comparator must match the bit width of the comparison target (32 bits for a VID).
The reducer, implemented as an adder tree, aggregates these results into one value that populates the pointer array in data reshaping.
Since the comparator produces 1-bit outputs, the reducer requires an adder width of up to $\log n$, where $n$ is the number of concurrently processed elements.

The reshaper receives the sorted COO from the UPE and temporarily stores it in a buffer. It then uses multiple SCR units to determine the frequency of occurrences for the current target VIDs ($v \sim v+n-1$) within the COO.
Subsequently, the reshaper maintains two counters: one for the progress on target VIDs (i.e., the current $v$), and another for the number of COO elements already consumed. Whenever a target VID meets a COO element with a value strictly larger than itself, the target VID is marked as completed, and $v$ is advanced to the next VID by updating the target counter accordingly.
Meanwhile, any COO element with VID smaller than $v$ is marked as consumed, as it can no longer contribute to the remaining targets. On consumption, the reshaper fetches the next COO segment into the buffer and reiterates the counting steps until all COO segments have been processed.

\enlargethispage{10pt}

For the reindexer, the hardware design is similar, except that the reducer adopts a filter tree (OR gates) instead of an adder tree.
In this case, the inputs represent mapping elements from the SRAM bank and the new VID to be reindexed.
As it is required for the reindexer to return the actual value (the reindexed VID), together with an indication of a search hit, the filter tree's bit width must match that of each element being filtered plus one (32+1 bits for a VID).
Figure \ref{fig:over_SCR_kernel_3} further shows the operation of the reindexer with its associated SCRs and SRAM bank. The reindexer maintains a counter to track the number of mappings completed so far. Two arrays are stored in the SRAM bank: one for original VIDs and another for renumbered VIDs (as described in Section \ref{sec:overview-hardware}).
For each index pair in these arrays, the SCR checks whether the original VID exists. If it does, the SCR returns the renumbered VID. If not, the reindexer increments the counter, assigns it as the new VID, and stores the input target and the counter value as a new mapping pair in the SRAM. This design ensures high-performance reshaping and subgraph reindexing.

\begin{figure*}
  \centering
  \vspace{-10pt}
  \includegraphics[width=\linewidth]{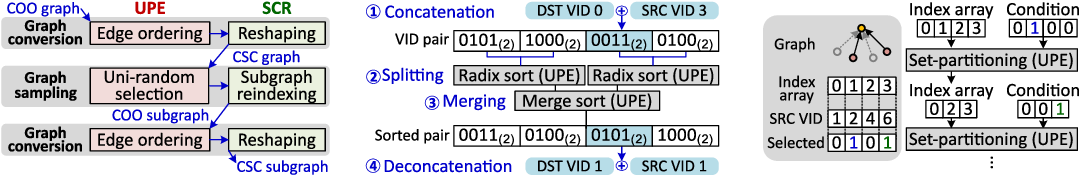}
  \begin{minipage}{0.28\linewidth}
    \vspace{7pt}
    \caption{End-to-end workflow.}\label{fig:impl_workflow}
    \vspace{0pt}
  \end{minipage}
  \begin{minipage}{0.38\linewidth}
    \centering
    \vspace{7pt}
    \caption{Workflow for edge ordering.}\label{fig:impl_UPE_edge_ordering}
    \vspace{0pt}
  \end{minipage}
  \begin{minipage}{0.32\linewidth}
    \centering
    \vspace{7pt}
    \caption{Graph sampling.}\label{fig:impl_graph_sampling}
    \vspace{0pt}
  \end{minipage}
  \vspace{-20pt}
\end{figure*}


\section{Details of End-to-End Operations}\label{sec:implementation}
\subsection{End-to-End Workflow and Dataflow}\label{subsec:e2e-workflow}

Figure \ref{fig:impl_workflow} shows the end-to-end GNN preprocessing workflow fully automated in hardware. It begins with the COO-to-CSC conversion, initiated by the UPE controller through edge ordering. The edge ordering workflow comprises four steps: concatenation, splitting, merging, and deconcatenation, all executed in parallel by multiple UPEs. Details regarding these steps will be explained shortly. After sorting the COO, the SCR controller (reshaper) generates the pointer array using multiple SCRs, as described in Section \ref{sec:impl-scr}. Independent set-counting allows each pointer to be processed in parallel.
After data reshaping, the source nodes associated with each pointer (destination VID) are identified. The UPE controller then begins unique random selection across source nodes for each pointer, using the same UPEs previously used for edge ordering.
This sequential use of shared hardware ensures no UPE remains idle, optimizing FPGA resource usage.
Meanwhile, the sampled nodes are managed by the SCR controller (reindexer). As described in Section \ref{sec:motiv-critical-remove}, the reindexer checks mapping information stored in the SRAM using set-counting and finalizes the sampled graph by renumbering.

\begin{figure}[b]
  \vspace{-10pt}
  \centering
  \footnotesize
  \removelatexerror
  \begin{algorithm}[H]
    \setstretch{0.9}
    \Fn{Merge($A$ and $B$: sorted edge arrays of length $n$) $\blacktriangleright$  $C$: the merge array}{
      $a, b, c = w/2, w/2, 0$   \tcp*{current index of $A, B, C$}\label{line:3-1-1}
      $buf[0:w/2], buf[w/2:w] = A[0:a], B[0:b]$ \;\label{line:3-1-2}
      \While{$a<n$ and $b<n$}{
        UPE\_sort($buf$) \tcp*{Sort the buffer}\label{line:3-2}
        $C[c:c+w/2] = buf[0:w/2]$; $c = c + w/2$ \;\label{line:3-3}
        \textbf{let} $is\_a = A[a] < B[b]$ \;\label{line:3-4-1}
        $buf[0:w/2] = is\_a ? A[a:a+w/2] : B[b:b+w/2]$\;%
        \leIf{$is\_a$}{$a = a+w/2$}{$b = b+w/2$}\label{line:3-4-2}
      }
    }
    \caption{\small Merge sorting using UPE.}\label{alg:ordering}
  \end{algorithm}
  \vspace{-15pt}
\end{figure}



\noindent \textbf{Workflow for edge ordering.}
While UPE has no limit on the number of elements it can process simultaneously, increasing the element count deepens the prefix-sum and relocation logic, degrading performance. To mitigate this, the UPE controller splits the COO format into smaller chunks matching the UPE's processing capacity, referred to as the \emph{UPE width}. It then performs radix sort on these chunks. In our FPGA implementation using VPK180 \cite{AMD23VPK180}, UPEs can be configured up to 240 instances, each with a width of 64 elements. However, the number of UPEs and their width can scale further depending on the size of the underlying FPGA.

\enlargethispage{10pt}


Figure \ref{fig:impl_UPE_edge_ordering} shows the UPE controller's workflow for edge ordering. Since the COO represents edges as pairs of destination VIDs and source VIDs, the UPE controller first concatenates these VIDs for each pair. It then splits the concatenated pairs into chunks based on the UPE width. Each UPE processes a chunk independently, performing radix sort (Section \ref{subsec:impl-upe}), resulting in locally sorted COO chunks.
To produce a fully sorted COO, the UPE controller employs UPEs to merge these locally sorted segments.
By leveraging the UPEs to find the minimum values among the sorted chunks simultaneously, the UPE controller eventually produces a globally sorted COO.
Algorithm \ref{alg:ordering} describes these steps in detail.
The merge process starts by reading the smallest $w/2$ elements from each input array, where $w$ is the UPE width (lines {\ref{line:3-1-1}}-{\ref{line:3-1-2}}), and sorting them (line {\ref{line:3-2}}). Since these are the smallest in both arrays, the first $w/2$ elements of the sorted buffer are returned (line {\ref{line:3-3}}). The buffer still holds $w/2$ elements, so the scheduler reads only $w/2$ new elements from either array $A$ or $B$. To decide which array to read, the scheduler compares the first elements in $A$ and $B$, then copies $w/2$ elements from the array with the smaller first element (lines {\ref{line:3-4-1}}-{\ref{line:3-4-2}}). By repeating this process, arrays $A$ and $B$ can be merged at a rate of $w/2$ elements per cycle.
Finally, the UPE controller deconcatenates the sorted elements back into pairs of destination and source VIDs to complete edge ordering.

\noindent\textbf{Control-path of graph sampling.}
The UPE controller samples graphs by selecting $k$ nodes from the source nodes, ensuring both uniqueness and randomness.
Note that for node-wise sampling methods \cite{hamilton2017inductive,jianfei2018stochastic}, the controller is called for every neighbor array. To support layer-wise sampling methods \cite{ying2018graph,chen2018fastgcn,wenbing2018adaptive,difan2019layerdependent}, all neighbor node arrays of a layer are first aggregated together as a single array; the UPE controller is then called for the aggregated array, selecting $k$ nodes per layer.
This flexibility can effectively support a wide range of popular GNN models.
Figure~\mbox{\ref{fig:impl_graph_sampling}} illustrates a simplified version of this process.
First, the controller creates an index array whose length equals the number of source nodes, each element corresponds to the index of the source node.
It then repeats the random selection procedure $k$ times.
Each iteration draws a new random index from the unsampled set to create a one-hot condition for that index, and let the UPEs run set-partitioning to extract the chosen element in a single cycle.
The UPE controller maintains a bitmap and updates it whenever a new index is extracted by marking the extracted index as sampled, preventing reselection, and thus guaranteeing uniqueness.
After $k$ indices are selected, the controller builds a condition array from
the bitmap and applies set-partitioning once more on the original source node array to extract the $k$-sampled neighborhood.


\subsection{Dynamic Reconfiguration and Software}\label{sec:dyn-reconfig}
AutoGNN sustains high performance even as graph characteristics shift at runtime by dynamically reconfiguring the accelerator.
To remove the long synthesis latency from the critical path, we do not synthesize hardware at runtime, but instead select among a small set of pre-compiled bitstreams.

\noindent\textbf{Bitstream generation.}
Both UPE and SCR are parameterizable in their count and width.
We therefore pre-compile a series of bitstreams that fit the target FPGA. 
Because both hardware are most efficient when configured with widths that are a power of two, we start from a bitstream consisting of a single large UPE (and SCR), 
and iteratively halve the width and double the instance count, generating the corresponding RTL and compiling them offline using Vivado. 
On our VMK180 board, this yields ten UPE variants and ten SCR variants, thus twenty kernel bitstreams in total.
Rather than compiling every UPE $\times$ SCR combination, AutoGNN partitions the device into two reconfigurable regions with a fixed area split of 70:30, which was effective in our evaluation (Section~\mbox{\ref{sec:eval-analysis}}).
The static partitioning allows us to pre-compile ten UPE bitstreams and ten SCR bitstreams separately.
At boot, all twenty bitstreams (50 MB each, 1 GB total) are staged in the internal DRAM of AutoGNN.
This confines the reconfiguration cost to bitstream loading and partial reconfiguration.
To this end, the reconfiguration process takes $\sim$ 230 ms, including 3 ms to load the bitstream from DRAM and 225 ms for FPGA reconfiguration through the Xilinx ICAP IP~\mbox{\cite{hwicap}} operating at 100 MHz.
Note that, because UPE and SCR reside in separate reconfigurable regions, only the region that needs to change could be reprogrammed, roughly halving the reconfiguration overhead. 

\begin{table}[b]
  \vspace{-10pt}
  \centering
  \setlength\tabcolsep{6pt}
  \resizebox{\linewidth}{!}{
    \setstretch{0.95}
    \small
    \begin{tabular}{|c|c|}
      \hline
      \rowcolor{gray!20} \textbf{Task} & \textbf{Cost function} \\
      \hline
      \multirow{2}{*}{\textbf{Edge ordering}} & $m = \log_2\left(e / w_{upe}\right) - 1 $ \\
                                              & $\textstyle \text{cycle}_{Ordering} = \left( \frac{2 \times m \times e}{n_{upe} \times w_{upe}} \right)$ \\
      \hline
      \textbf{Unique}                         & $s = b \times k^{l+1}-1$ \\
      \textbf{random selection}               & $\text{cycle}_{Selecting} = s / n_{upe}$ \\
      \hline
      \textbf{Data reshaping}                 & $\text{cycle}_{Reshaping} = \max\left(\frac{n}{n_{scr}}, \frac{e}{w_{scr}}\right) $ \\
      \hline
    \end{tabular}
  }
  \vspace{3pt}
  \caption{Cost functions of GNN preprocessing tasks.}\label{tbl:coo2csr_upe}
  \vspace{-14pt}
\end{table}

\begin{figure*}
  \vspace{-5pt}
  \centering
  \begin{minipage}[t]{0.68\linewidth}
    \vspace{0pt}
    \resizebox{1\linewidth}{!}{
      \setlength{\tabcolsep}{1.5pt}
      \hspace{-10pt}
      \begin{tabular}{@{}cccccccccccc@{}}
        \toprule
                                 & \multicolumn{2}{c}{Category} & \#Edges  & \#Nodes  & Deg    &                          & \multicolumn{2}{c}{Category} & \#Edges  & \#Nodes  & Deg    \\
        \midrule%
        \cellcolor[HTML]{fee7d1} & Physics \cite{wang2019deep}  & (\textbf{PH})     & 495K     & 34.5K    & 14.4   & \cellcolor[HTML]{e6dfee} & Reddit2 \cite{fey2019fast}  & (\textbf{RD})    & 23.2M    & 233K     & 99.6   \\ \cline{2-6} \cline{8-12}
        \cellcolor[HTML]{fee7d1} & arxiv \cite{hu2020ogb}    & (\textbf{AX})     & 1.16M    & 169K     & 6.84   & \cellcolor[HTML]{e6dfee} & StackOver \cite{snapnets} & (\textbf{SO})    & 63.5M    & 6.02M    & 10.5   \\ \cline{2-6} \cline{8-12}
        \multirow{-3}{*}{\cellcolor[HTML]{fee7d1}\textbf{\begin{tabular}[c]{@{}c@{}}Citation\\Network\end{tabular}}}%
                                 & collab \cite{hu2020ogb}  & (\textbf{CL})     & 2.36M    & 236K     & 10.0   &
        \multirow{-3}{*}{\cellcolor[HTML]{e6dfee}\textbf{\begin{tabular}[c]{@{}c@{}}Social\\Network\end{tabular}}}        & Journal~\cite{snapnets}  & (\textbf{JR})    & 69.0M    & 4.85M    & 14.2   \\
        \midrule%
        \cellcolor[HTML]{dbf0db} & Yelp \cite{wang2019deep}    & (\textbf{YL})     & 6.81M    & 46.0K    & 148    & \cellcolor[HTML]{fcfcd1} & Amazon \cite{hu2020ogb}   & (\textbf{AM})    & 123M     & 2.45M    & 50.5   \\ \cline{2-6} \cline{8-12}
        \cellcolor[HTML]{dbf0db} & Fraud \cite{wang2019deep}   & (\textbf{FR})     & 7.13M    & 11.9K    & 597    & \cellcolor[HTML]{fcfcd1} & Taobao \cite{taobaodataset}    & (\textbf{TB})    & 400M     & 230K     & 1744   \\ \cline{2-6} \cline{8-12}
        \multirow{-3}{*}{\cellcolor[HTML]{dbf0db}\textbf{\begin{tabular}[c]{@{}c@{}}Interaction\\Network\end{tabular}}}%
                                 & Movie \cite{wang2019deep}   & (\textbf{MV})     & 11.3M    & 3.71K    & 3052   &
        \multirow{-3}{*}{\cellcolor[HTML]{fcfcd1}\textbf{\begin{tabular}[c]{@{}c@{}}E-commerce\\Network\end{tabular}}}    &           &                  &          &          &        \\
        \bottomrule
      \end{tabular}
    }
    \vspace{4pt}
  \captionof{table}{Important characteristics of dataset.}
  \label{tbl:datasets_summary}
  \end{minipage}%
  \begin{minipage}[t]{0.01\linewidth}
    \vspace{0pt}
    \hspace{1pt}
  \end{minipage}%
  \begin{minipage}[t]{0.31\linewidth}%
    \vspace{0pt}
    \center
    \includegraphics[width=1.\linewidth]{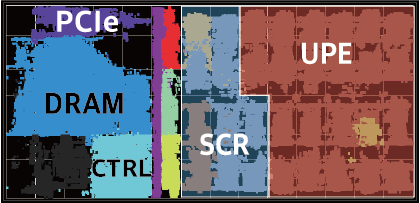}
    \vspace{-10pt}
    \caption{Floorplan.}\label{fig:result_floorplan}
  \end{minipage}%
  \vspace{-10pt}
\end{figure*}

\noindent\textbf{Cost function.}
At runtime, the host collects light-weight graph metadata (e.g., the number of nodes $n$ and edges $e$) and GNN hyperparameters (e.g., the number of layers $l$, the max sample count $k$, and the batch size $b$).
A cost function then estimates the end-to-end latency for each pre-compiled bitstreams to find the best hardware configuration for the specified input graph and GNN model.
Specifically, the cost function evaluates three analytic models, each for ordering, selection, and reshaping as shown in Table~\mbox{\ref{tbl:coo2csr_upe}}.
The cost functions are parameterized by the hardware and workload (input graph and GNN model) configuration.
The hardware parameters includes the number and width of UPE/SCR engines ($n_{\text{upe}}, w_{\text{upe}}, n_{\text{scr}}, w_{\text{scr}}$, respectively), while the workload parameters include $n$, $e$, $l$, $k$, and $b$.

For edge ordering, each UPE processes its assigned chunk independently. Therefore, the runtime is determined by the dividing total work, which is equal to the number of edges ($e$) multiplied by the number of merging rounds ($m$),  by the aggregated UPE throughput, which scales proportional to the throughput of each UPE ($w_{upe}/2$) and the number of UPEs ($n_{upe}$).
For uni-random selection, each UPE selects one node per step, yielding $n_{upe}$ nodes per step across all UPEs.
The runtime is therefore estimated by dividing the total number of nodes selected ($s$) by $n_{upe}$, where $s$ is calculated as the product of the batch size ($b$) and $k$ nodes selected per layer ($l$).
For \emph{data reshaping}, SCRs process the COO array in parallel to build the CSC pointer array. Time is bounded by the maximum of two terms: the COO-side term, which is proportional to the number of edges ($e$) divided by the SCR width ($w_{scr}$), and the CSC-side term, which is proportional the number of nodes ($n$) in the pointer array divided by the number of SCRs ($n_{scr}$).
Evaluating the cost function consists of two steps: i) evaluating the cost-related graph parameters, which is handled by AutoGNN during graph conversion, and ii) scoring the costs for all available bitstreams, which is done by the host library.
In our evaluation, the cost computation took less than 0.1 ms, which is under 0.1\% of the end-to-end latency.


\enlargethispage{10pt}

\begin{figure}[b]
  \vspace{-8pt}
  \begin{minipage}[t]{0.56\linewidth}
    \strut\vspace*{-\baselineskip}\newline
    \resizebox{1\linewidth}{!}{
      \setlength{\tabcolsep}{1.5pt}
      \begin{tabular}{|
        >{\columncolor[HTML]{D9D9D9}}c |c|c|}
        \hline
        \cellcolor[HTML]{D9D9D9}                                                                              & CPU              & \begin{tabular}[c]{@{}c@{}}Xeon   128-core\\ 512GB, DDR5\end{tabular} \\ \cline{2-3}
        \cellcolor[HTML]{D9D9D9}                                                                              & GPU              & RTX   3090                                                            \\ \cline{2-3}
        \multirow{-4}{*}{\cellcolor[HTML]{D9D9D9}\textbf{\begin{tabular}[c]{@{}c@{}}Evaluation\\ System\end{tabular}}}   & FPGA             & \begin{tabular}[c]{@{}c@{}}VPK180\\ (4.1M LUT)\end{tabular}           \\ \hline
        \cellcolor[HTML]{D9D9D9}                                                                              & Framework        & DGL   2.3.0                                                           \\ \cline{2-3}
        \cellcolor[HTML]{D9D9D9}                                                                              & GNN Model           & \begin{tabular}[c]{@{}c@{}}2-layer\\ GraphSAGE\end{tabular}                                                             \\ \cline{2-3}
        \cellcolor[HTML]{D9D9D9}                                                                              & Selecting $k$           & 10                                                             \\ \cline{2-3}
        \multirow{-5}{*}{\cellcolor[HTML]{D9D9D9}\textbf{\begin{tabular}[c]{@{}c@{}}Software\\ Configuration\end{tabular}}} & Inf. Nodes   & 3000                                                                  \\ \hline
        \cellcolor[HTML]{D9D9D9}                                                                              & SCR Resource & 30\%                                                                  \\ \cline{2-3}
        \cellcolor[HTML]{D9D9D9}                                                                              & SCR Slots    & 1                                                                     \\ \cline{2-3}
        \multirow{-3}{*}{\cellcolor[HTML]{D9D9D9}\textbf{\begin{tabular}[c]{@{}c@{}}Hardware\\ Configuration\end{tabular}}}  & UPE Width        & 64                                                                    \\ \hline
        \end{tabular}
    }
    \vspace{5pt}
    \captionof{table}{Evaluation setup.}
    \label{tbl:device_parameters}
  \end{minipage}
  \begin{minipage}[t]{0.42\linewidth}
    \strut\vspace*{-\baselineskip}\newline
    \resizebox{1\linewidth}{!}{
      \setlength{\tabcolsep}{1.5pt}
      \begin{tabular}{cc}
        \textbf{Operation} & \textbf{Algorithm} \\
        \specialrule{.2em}{.1em}{.1em}
        Ordering & \begin{tabular}[c]{@{}c@{}}Radix\\ Sort \cite{andrey20sort}\end{tabular} \\
        \hline
        Reshaping & \begin{tabular}[c]{@{}c@{}}Histogram\\ Hashing \cite{Daniel23Hash}\end{tabular} \\
        \hline
        Selecting & \begin{tabular}[c]{@{}c@{}}Reservoir\\ Sampling \cite{vitter1985random}\end{tabular} \\
        \hline
        Reindexing & \begin{tabular}[c]{@{}c@{}}Histogram\\ Hashing \cite{Daniel23Hash}\end{tabular} \\
      \end{tabular}
    }
    \vspace{17pt}
    \captionof{table}{Algorithms.}
    \label{tbl:operations_algorithms}
  \end{minipage}
  \vspace{-15pt}
\end{figure}

\begin{figure*}
  \vspace{-10pt}
  \centering
  \includegraphics[width=\linewidth]{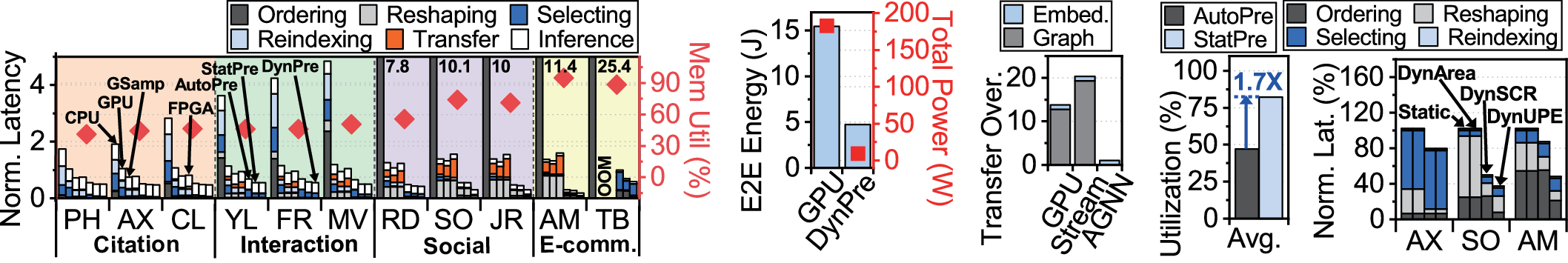}
  \begin{minipage}{0.48\linewidth}%
    \vspace{5pt}
    \caption{End-to-end latency.}\label{fig:result_overall}
  \end{minipage}%
  \hspace{7pt}
  \begin{minipage}{0.06\linewidth}%
    \vspace{0pt}
    \caption{Power.}\label{fig:result_power}
  \end{minipage}%
  \hspace{25pt}
  \begin{minipage}{0.07\linewidth}
    \vspace{0pt}
    \caption{Transfer.}\label{fig:result_transfer}
  \end{minipage}%
  \hspace{17pt}
    \begin{minipage}{0.105\linewidth}%
    \vspace{0pt}
    \caption{LUT utilization.}\label{fig:result_lut_utilization}
  \end{minipage}%
  \hspace{8pt}
  \begin{minipage}{0.145\linewidth}%
    \vspace{0pt}
    \caption{Dynamic \mbox{reconfiguration.}}\label{fig:result_reconfig_mix}
  \end{minipage}%
  \vspace{-20pt}
\end{figure*}

\noindent \textbf{Software architecture.}
We modified the deep graph library (DGL) to integrate AutoGNN into GNN inference services using commodity GPUs (or hardware accelerators). All interfaces remain consistent with DGL, with additional functions for graph management, such as \texttt{uploadgraph()}, which is similar to \texttt{updategraph()}. To maintain compatibility with DGL interfaces, the implementation details for managing AutoGNN are abstracted behind two software components: AutoGNN's user-level library (AGNN-lib) and kernel driver (AGNN-drv). AGNN-lib is responsible for i) managing graph I/O, ii) determining hardware reconfiguration, and iii) interacting with AutoGNN for preprocessing tasks.

Although graph datasets are smaller than their embeddings, they are large enough to be scattered in user memory (e.g., GB-scale). To handle such graphs, AGNN-drv utilizes AutoGNN's DMA-main interface through \texttt{pci\_ioremap\_bar()} \cite{pcieioremap} and processes them using a scatter-gather list \cite{scattergather}. AGNN-drv creates a PCIe descriptor for the scatter-gather list based on the given graph data and writes the descriptor address to DMA-main. AutoGNN then retrieves the graph dataset by referencing the descriptor and preprocesses it.
Unlike the GPU, which must deallocate the graph datasets during the model inference process, AutoGNN can store the previous graph data within device memory. This enables AutoGNN to only read the updated portions of the graph from the host, minimizing data transfer overheads.
Instead, AutoGNN requires an additional data movement to transfer the preprocessed graph to the GPU. However, since the preprocessed graph is already sampled, its size is substantially smaller than the original graph, making this overhead negligible. In typical GNN scenarios, this transfer latency is approximately 2.8 ms, accounting for less than 1\% of the total end-to-end GNN latency.

AGNN-lib evaluates the cost function if the current hardware configuration is suboptimal due to changes in the graph.
If the latency exceeds the threshold, AGNN-lib selects the optimal pair of hardware profiles (one for UPE, one for SCR) and sends its key to AutoGNN. The AutoGNN's hardware shell, specifically the FPP controller, uses this key to locate the bitstream's address in the internal DRAM memory and DMA it to the ICAP IP to complete the reconfiguration.

\section{Evaluation}\label{sec:evaluation}

\noindent\textbf{Prototype.}
We constructed an AutoGNN hardware prototype implementing modules in RTL and synthesizing them onto a 7nm Xilinx VPK180 FPGA~\cite{AMD23VPK180}.
Figure~\ref{fig:result_floorplan} shows the example floorplan of our hardware, configured with eight SCR modules and 32 UPE modules respectively.
It is connected to a host system with a 128-core Xeon CPU and an RTX 3090 GPU, also used for the compared CPU and GPU-based systems.

\noindent\textbf{Tested model and workloads.} We employed a 2-layer GraphSAGE model~\cite{Liu2021SamplingMF}, which selects 10 neighboring samples for each node during preprocessing (i.e., $k$ is set to 10) and uses the resulting 2-hop subgraphs for inference. We selected 11 graph datasets from OGB~\cite{hu2020ogb}, DGL~\cite{wang2019deep}, and PyG~\cite{fey2019fast}. These datasets can be categorized into four groups based on their domains and characteristics: (i) {\em Citation networks} represent relationships between papers by modeling papers as nodes and citations as edges. These networks typically have small sizes and degrees. (ii) {\em Interaction networks} express relationships between movies or restaurants by modeling them as nodes and their reviews as edges, resulting in high connectivity. (iii) {\em Social networks} are graphs that represent relationships between individuals or organizations, which are typically large and exhibit medium connectivity. (iv) {\em E-commerce networks} are graphs that model customers or products as nodes and purchases or searches as edges, which are typically large. Table~\ref{tbl:datasets_summary} summarizes important characteristics of the datasets. 

\enlargethispage{20pt}

\begin{figure}[b]
  \vspace{-10pt}
  \begin{subfigure}{0.49\linewidth}
    \includegraphics[width=\linewidth]{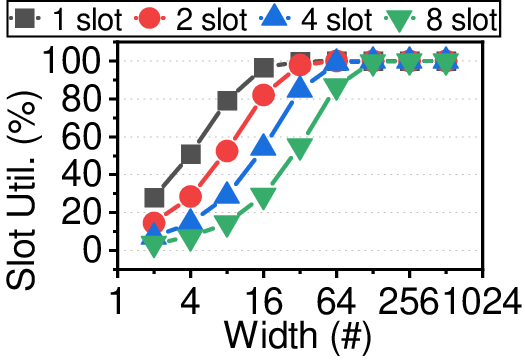}
    \vspace{-13pt}
    \caption{SCR width (AX).}\label{fig:result_sens_slot_util}
  \end{subfigure}
  \begin{subfigure}{0.49\linewidth}
    \includegraphics[width=\linewidth]{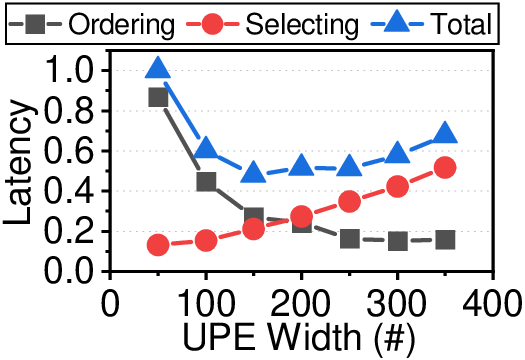}
    \vspace{-13pt}
    \caption{UPE width (AM).}\label{fig:result_sens_upe}
  \end{subfigure}
  \vspace{-5pt}
  \caption{Optimal hardware configuration.}\label{fig:result_sens_optimal}
  \vspace{-15pt}
\end{figure}

\noindent\textbf{Compared systems and configurations.}
We compared seven systems: four baselines and three versions of AutoGNN. Among the baselines, \texttt{CPU} and \texttt{GPU} perform preprocessing on the CPU and GPU, respectively.
Both use the DGL~\cite{wang2019deep} framework, and the specific algorithms for each preprocessing step are summarized in Table~\ref{tbl:operations_algorithms}.
We also evaluated two configurations that accelerate the graph sampling process.
\texttt{GSamp}~\cite{gong2023gsampler} deploys matrix-centric APIs compiled via a data-flow IR with fusion and super-batching.
\texttt{FPGA}~\cite{gui2024fpga} utilizes a stream-based sampler logic backed by HBM. Since the accelerator implements sampling only, we run graph conversion on the GPU.
The other three systems are our AutoGNN variants, all of which execute end-to-end GNN preprocessing on the FPGA. They differ in how the UPE region is organized for the two stages (ordering and selection) and whether the hardware kernels are reconfigurable.
In \texttt{AutoPre}, the UPE region is statically split into two fixed sub-engines: an ordering-only UPE and a selection-only UPE, each provisioned with equal LUT budgets. This design deliberately forgoes the UPE's unification capability by keeping both stage-specific datapaths, but the two stages still execute serially due to dependencies.
\texttt{StatPre} utilizes the whole UPE region in a time-multiplexed manner across ordering and selection. Reusing the UPE, which utilizes most (70\%) of the FPGA resources, improves overall resource utilization compared to \texttt{AutoPre}.
\texttt{DynPre} additionally leverages partial reconfiguration to adapt the UPE and SCR components to the target dataset at runtime.
Unless otherwise noted, the hardware settings of \texttt{AutoPre} and \texttt{StatPre} are fixed and tuned for the MV dataset (an intermediate-sized graph) for best average performance. After preprocessing, all systems perform GNN inference on the GPU. Table~\ref{tbl:device_parameters} lists the key software and hardware configurations.

\enlargethispage{10pt}

\subsection{Overall Performance}\label{sec:eval-perf}

\noindent\textbf{End-to-end latency.}
Figure~\ref{fig:result_overall} compares end-to-end GNN inference latency normalized to \texttt{GPU}.
Compared to \texttt{CPU}, the \texttt{GPU}, \texttt{GSamp}, \texttt{FPGA}, \texttt{AutoPre}, \texttt{StatPre}, and \texttt{DynPre} systems reduce overall latency by 3.4$\times$, 4.5$\times$, 4.1$\times$, 7.3$\times$, 8.4$\times$, and 9.0$\times$, respectively.
\texttt{GPU} delivers a large speedup for edge ordering by 3421$\times$ compared to \texttt{CPU} thanks to its massive parallelism. However, other GNN preprocessing operations exhibit heavy atomic operations which limit GPU performance. Consequently, the end-to-end speedup of \texttt{GPU} over \texttt{CPU} averages only 3.4$\times$ on average.
\texttt{GSamp} and \texttt{FPGA} further accelerate sampling by 7.5$\times$ and 12$\times$, respectively.
Unlike \texttt{GSamp}, because \texttt{FPGA} implements sampling only, other stages of the GNN, including graph conversion, must run on the GPU. This necessitates the transfer of the full graph between the GPU and the FPGA. The resulting data movement accounts for 24.7\% of the end-to-end latency, on average.
In contrast, \texttt{AutoPre} automates all preprocessing to the FPGA, so only the graph updates are exchanged with the host. This reduces the transfer overhead by 19.2$\times$ and 35.4$\times$, on average, compared to \texttt{GPU} and \texttt{FPGA}.
Further analysis regarding the transfer overhead is explained through Figure~\ref{fig:result_transfer}.
Additionally, \texttt{AutoPre} utilizes SCRs to accelerate reshaping and reindexing, which require atomic operations. To this end, although \texttt{AutoPre}'s per-stage throughput for ordering and selecting can be lower than \texttt{FPGA}, \texttt{AutoPre} can still provide 1.9$\times$ performance boost, on average.
\texttt{StatPre} improves upon \texttt{AutoPre} by unifying the UPE region; the same UPE module is time-multiplexed across ordering and selection, which raises overall resource utilization. This shortens the end-to-end latency by 14\%, on average.
\texttt{DynPre} enables partial reconfiguration within the UPE/SCR region, adapting the hardware to each dataset at runtime. This further reduces the end-to-end latency by 21.6\% on average. The gains of \texttt{DynPre} are most pronounced for large or low-degree graphs, which differ substantially from MV, where the preprocessing time drops by 53.6\%.


\begin{figure}[b]
  \vspace{-10pt}
  \includegraphics[width=0.495\linewidth]{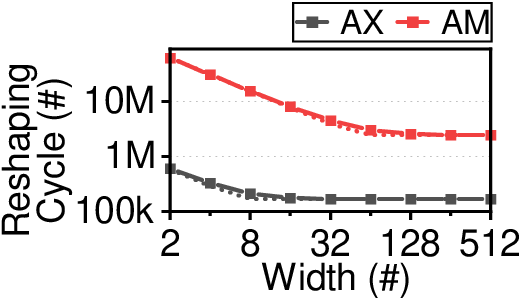}
  \includegraphics[width=0.495\linewidth]{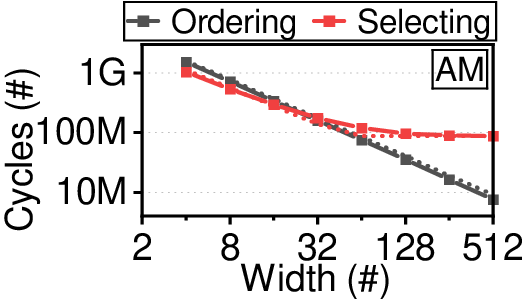}
  \vspace{-4pt}
  \caption{Accuracy of the cost model.}\label{fig:result_cost}
  \vspace{-32pt}
  \begin{subfigure}{1\linewidth}
    \renewcommand*{\arraystretch}{0.3}
    \begin{tabularx}{\textwidth}{
      p{\dimexpr.495\linewidth-1.33\tabcolsep}
      p{\dimexpr.495\linewidth-1.33\tabcolsep}
      }
      \caption{Accuracy of SCR cycle.}\label{fig:result-cost-scr} &
      \caption{Accuracy of UPE cycle.}\label{fig:result-cost-upe}
    \end{tabularx}
  \end{subfigure}
  \vspace{-4pt}
\end{figure}

\noindent\textbf{Memory bandwidth utilization.}
The right y-axis in Figure~\ref{fig:result_overall} reports the memory bandwidth utilization of \texttt{DynPre}.
AutoGNN sustains an average utilization of 59.8\%, noticeably higher than the GPU baseline (30.3\% on average).
This gap stems from AutoGNN's specialized datapaths. UPE and SCR implement prefix-sum and reduction logic with pipelined adders and adder trees, respectively, replacing serialized atomic operations with single-pass kernels.
The memory utilization increases with graph size. Larger graphs tend to have long latencies for reshaping, which processes sorted COO segments sequentially; with SCR, each COO segment is consumed in a single cycle, allowing the SCR to fully saturate the memory interface. Consequently, we observe an average utilization of 91.6\% on e-commerce graphs.


\noindent\textbf{Power and energy.}
Figure~\ref{fig:result_power} compares AutoGNN (\texttt{DynPre}) and \texttt{GPU} in terms of power and total energy.
During preprocessing, \texttt{DynPre} draws only 9.3W on the FPGA, whereas \texttt{GPU} dissipates 183W for the same workload, yielding a 19.7$\times$ lower power draw for preprocessing.
When considering end-to-end GNN inference, both configurations execute the GNN model on the GPU, which narrows the energy gap. Even so, thanks to the latency reduction of \texttt{DynPre}, the total energy consumption is on average 3.3$\times$ lower than \texttt{GPU}.

\enlargethispage{10pt}


\subsection{Detailed Analysis}\label{sec:eval-analysis}

\noindent\textbf{Transfer overhead.}
Figure~\ref{fig:result_transfer} shows the average transfer overhead for \texttt{GPU}, \texttt{FPGA}, and \texttt{AutoPre}.
\texttt{AutoPre} reduces transfer overhead by 13.6$\times$ and 20$\times$ compared to \texttt{GPU} and \texttt{FPGA}, respectively.
This stems from AutoGNN's ability to perform end-to-end preprocessing entirely on the accelerator.
For \texttt{GPU}, the same device must handle both preprocessing and GNN model execution. Due to the lack of GPU's internal memory, the entire graph must be fetched from the host again before each preprocessing pass.
For \texttt{FPGA}, on top of the host-GPU transfers, an additional transfer is required to move the CSR-form of input graph produced by the GPU-side graph conversion. 
In contrast, \texttt{AutoPre} executes all preprocessing stages on a single accelerator and therefore only receives the incremental graph updates from the host. After preprocessing, the resulting sampled subgraph is sent to the GPU for model execution; however, this subgraph is much smaller than the original graph (1230$\times$ on average), so the residual transfer overhead is minimal (0.6\% of end-to-end latency).


\begin{figure}
  \vspace{-10pt}
  \includegraphics[width=\linewidth]{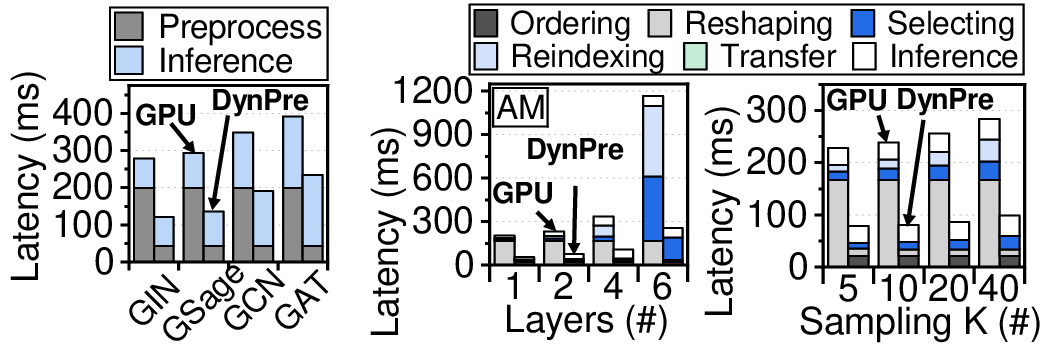}
  \vspace{0pt}
  \begin{subfigure}[t]{0.38\linewidth}
    \vspace{0pt}
    \centering
    \vspace{-15pt}
    \caption{GNN models.}\label{fig:result_sensitivity_gnn_model}
    \vspace{-5pt}
  \end{subfigure}
  \begin{subfigure}[t]{0.3\linewidth}
    \vspace{0pt}
    \centering
    \vspace{-15pt}
    \caption{Layer count.}\label{fig:result-sensitivity-layer}
    \vspace{-5pt}
  \end{subfigure}
  \begin{subfigure}[t]{0.3\linewidth}
    \vspace{0pt}
    \centering
    \vspace{-15pt}
    \caption{$k$-selection.}\label{fig:result-sensitivity-k}
    \vspace{-5pt}
  \end{subfigure}
  \vspace{-15pt}
  \caption{Diverse GNN model support.}\label{fig:result_sensitivity_model}
  \vspace{-18pt}
\end{figure}

\noindent\textbf{LUT utilization.}
Figure~\mbox{\ref{fig:result_lut_utilization}} compares the average LUT utilization of \texttt{AutoPre} and \texttt{StatPre}.
\texttt{AutoPre} forgoes UPE unification and statically partitions the UPE region into distinct accelerators. 
Because data dependencies force serial execution, the average LUT utilization is only 47\%.
\texttt{StatPre} improves utilization to 82.2\% (1.7$\times$ over \texttt{AutoPre}) by time-multiplexing the same UPE modules across ordering and selection.



\noindent\textbf{Analysis of hardware reconfiguration.}
To evaluate hardware reconfiguration impact on performance enhancement, we implement three versions of \texttt{DynPre} and compare them with \texttt{StatPre}. \texttt{DynArea} uses an optimal resource distribution between the SCR and UPE modules for each dataset.
Compared to \texttt{DynArea}, \texttt{DynSCR} optimally configures the SCR module by optimizing its width and slot count. Compared to \texttt{DynSCR}, \texttt{DynUPE} optimizes the configuration of the UPE module in terms of width and count.
In this evaluation, we use three graph datasets as representatives: AX, SO, and AM.

\enlargethispage{10pt}

Figure~\ref{fig:result_reconfig_mix} compares the preprocessing latency of \texttt{StatPre}, \texttt{DynArea}, \texttt{DynSCR}, and \texttt{DynUPE} for the three datasets; the latency values are normalized to \texttt{StatPre}.
Compared to \texttt{DynArea}, \texttt{StatPre} assigns 30\% of the total LUT resources to the SCR module for every dataset, showing that adjusting the balance between the SCR and UPE modules brings negligible performance benefits. This analysis leads to fixing the SCR and UPE module resource distribution at 30:70. Recall that this decision significantly reduces reconfiguration overhead (Section~\ref{sec:dyn-reconfig}).
Next, compared to \texttt{DynArea}, \texttt{DynSCR} reduces the preprocessing latency of AX, SO, and AM by 23\%, 51\%, and 15\%, respectively. This indicates that the optimal configuration of the SCR module for all three datasets is distinct from that in MV. Figure~\ref{fig:result_sens_slot_util} further depicts the slot utilization under varying widths and counts for AX. Basically, slot utilization increases as slot width increases, since the number of edges read per cycle also increases; however, it does not further increase beyond a certain slot width, since the number of nodes recognized per cycle is limited by the slot count. To this end, for AX, which has a small degree, it is more beneficial to increase the number of slots.
Lastly, \texttt{DynUPE} reduces the preprocessing latency of SO and AM by 13\% and 39\%, compared to \texttt{DynSCR}. Figure~\ref{fig:result_sens_upe} shows the times for ordering and selecting operations as well as the total time under varying UPE widths and numbers, which highlights how the optimal UPE configuration is determined.

\noindent\textbf{Accuracy of cost model.}
Figure~\ref{fig:result_cost} compares cycles consumed (solid lines) with cycles estimated by our cost model (dotted lines) during preprocessing. Figure~\ref{fig:result-cost-scr} shows consumed and estimated cycles in the SCR module under varying slot widths for AX and AM, with a fixed SCR slot count. Our model achieves 98\% accuracy and captures each dataset's saturation. Figure~\ref{fig:result-cost-upe} plots cycles for ordering and selecting in the UPE module with varying UPE widths for AM. Our model achieves 94\% accuracy and effectively captures saturation, enabling identification of an optimal UPE configuration.

\enlargethispage{10pt}


\begin{figure}
  \vspace{-10pt}
  \includegraphics[width=\linewidth]{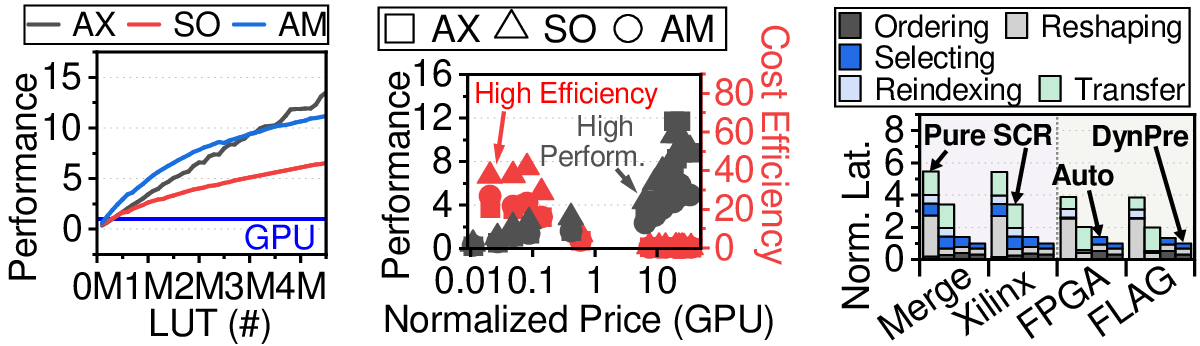}
  \begin{minipage}[t]{0.7\linewidth}%
    \vspace{0pt}
    \caption{Cost effectiveness.}\label{fig:eval-cost-effectiveness}
    \vspace{-30pt}
    \begin{subfigure}{1\linewidth}
      \renewcommand*{\arraystretch}{0.3}
      \begin{tabularx}{\textwidth}{
        p{\dimexpr.35\linewidth-1.33\tabcolsep}
        p{\dimexpr.65\linewidth-1.33\tabcolsep}
        }
        \caption{LUT.}\label{fig:eval-cost-effectiveness-lut.} &
        \caption{Price.}\label{fig:eval-cost-effectiveness-price.}
      \end{tabularx}
    \end{subfigure}
  \end{minipage}%
  \begin{minipage}[t]{0.3\linewidth}%
    \vspace{-18pt}
    \caption{Existing accelerators.}\label{fig:hpca-acc-sens}
  \end{minipage}%
  \vspace{-15pt}
\end{figure}

\noindent\textbf{Sensitivity on model parameters.}
Figure~\ref{fig:result_sensitivity_model} illustrates the end-to-end latency of GPU and DynPre with variations in the GNN model (Figure~\ref{fig:result_sensitivity_gnn_model}), the number of layers (Figure~\ref{fig:result-sensitivity-layer}), and the neighbor sampling parameter $k$ (Figure~\ref{fig:result-sensitivity-k}), for the AM dataset.
We analyzed four distinctive models -- GIN\cite{xiyuan2022how}, GraphSAGE\cite{hamilton2017inductive}, GCN\cite{kipf2016semi}, GAT\cite{velivckovic2017graph} -- ordered by computational intensity.
As the model complexity increases, the portion of GNN preprocessing decreases, narrowing AutoGNN's relative advantage. Nevertheless, even for the most demanding model (GAT), preprocessing still accounts for 51\% of the end-to-end latency, and \texttt{DynPre} delivers a 1.67$\times$ speedup over \texttt{GPU}.
AutoGNN can also support a wide range of GNN hyperparameters.
Increasing the number of layers from one to six increases the inference and sampling latency by 4.1$\times$ and 51.1$\times$, respectively. Consequently, the overall preprocessing overhead increases, boosting DynPre's relative speedup from 3.7$\times$ to 4.5$\times$.
Similarly, increasing $k$ amplifies the sampling latency by 2.5$\times$, and \texttt{DynPre}'s performance gain over GPU reaches 2.6$\times$ due to its streaming, contention-free sampler.

\noindent\textbf{Sensitivity on LUT and price.}
Figure~\ref{fig:eval-cost-effectiveness} evaluates the relative performance of \texttt{DynPre}, compared to \texttt{GPU}, while varying the total LUT count (Figure~\ref{fig:eval-cost-effectiveness-lut.}) and using different FPGA boards in a wide price range (Figure~\ref{fig:eval-cost-effectiveness-price.}).
As the LUT count increases from 400K to 4M, the relative performance rises from 1.9$\times$ to 9.6$\times$ on average. Given that the 3090 GPU has similar costs to a Xilinx FPGA with 400K LUTs~\cite{kintexfpga}, FPGA-based preprocessing achieves a 1.9$\times$ speedup at an equivalent cost.
We further analyzed the performance and cost effectiveness (the performance divided by the price) of AutoGNN on various FPGA boards.
On low-price FPGAs, AutoGNN shows moderate speedup against GPU (1.2$\times$ on average), but provides high cost effectiveness (21.8$\times$ on average).
On high-price FPGAs, AutoGNN's cost effectiveness falls; lying between 1.67$\times$ and 0.55$\times$, but the speedup rises to 7.6$\times$ on average.
To this end, AutoGNN's flexible hardware design enables consumers to choose the target FPGA based on their priority: if they want cost effectiveness, a low-end FPGA will suffice; if they lean toward compute power, a high-end FPGA provides greater performance than GPUs.


\begin{figure}
  \vspace{-10pt}
  \begin{subfigure}[t]{0.49\linewidth}
    \vspace{0pt}
    \centering
    \includegraphics[width=\linewidth]{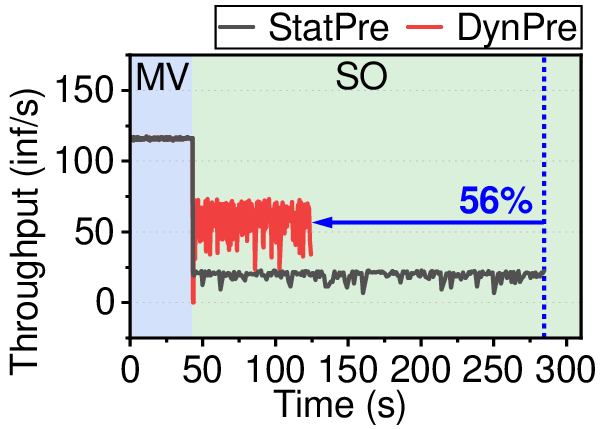}
    \vspace{-12pt}
    \caption{Time-series analysis.}\label{fig:result_timeline-g}
  \end{subfigure}
  \begin{subfigure}[t]{0.49\linewidth}
    \vspace{0pt}
    \centering
    \includegraphics[width=\linewidth]{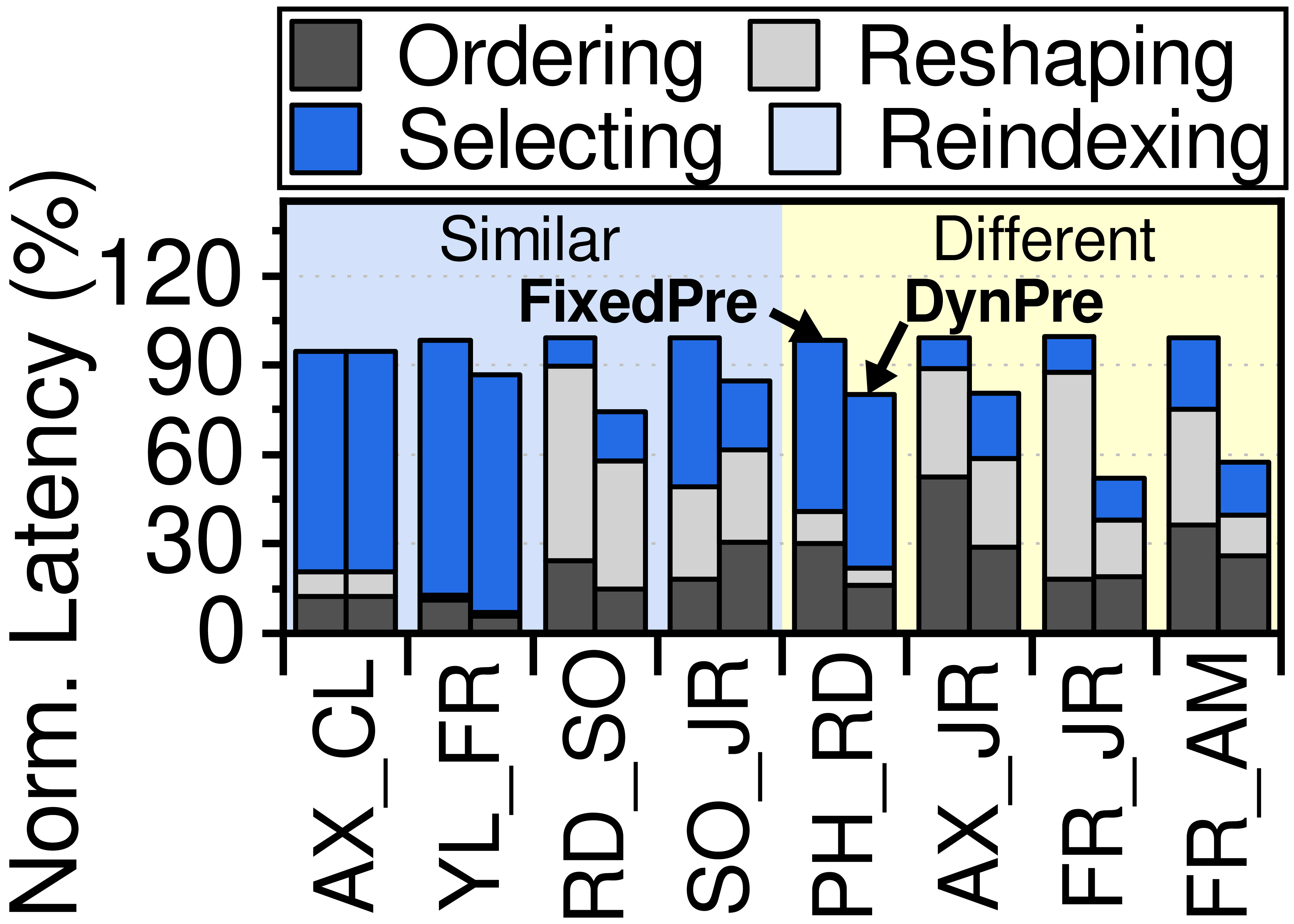}
    \vspace{-12pt}
    \caption{Various graph pairs.}\label{fig:result_mixed_timeline}
  \end{subfigure}
  \vspace{-6pt}
  \caption{Consecutive inference using diverse graphs.}\label{fig:result_timeline}
  \vspace{-15pt}
\end{figure}

\begin{figure}[b]
  \vspace{-10pt}
  \includegraphics[width=\linewidth]{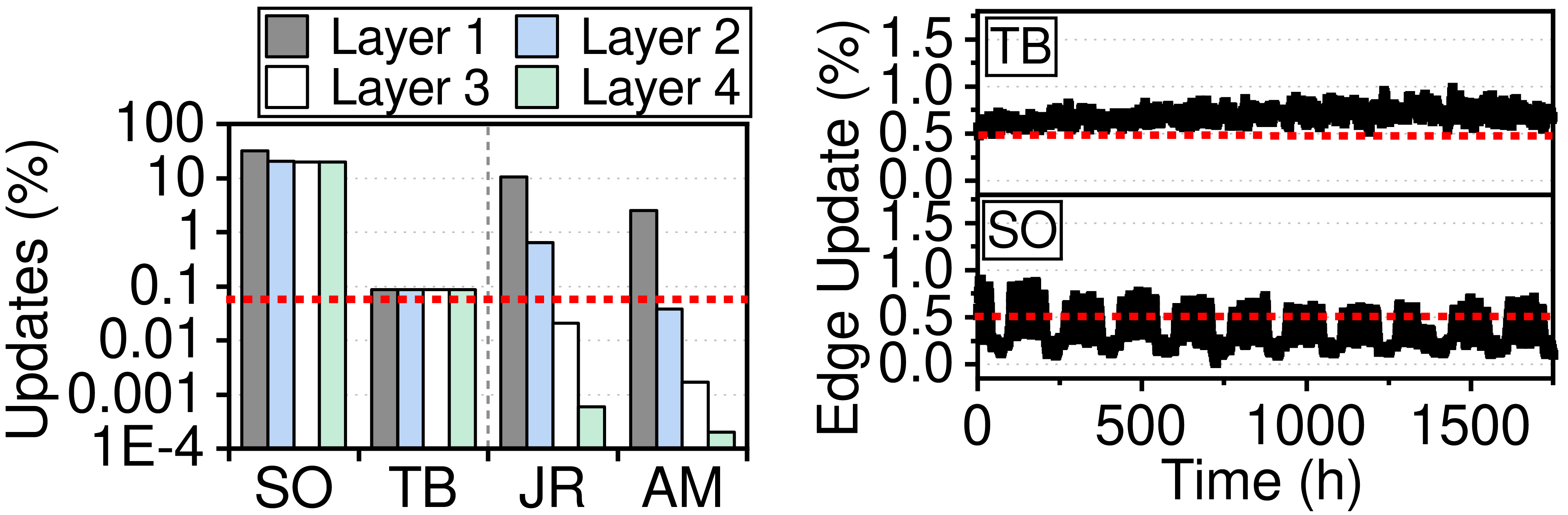}
  \begin{subfigure}[t]{0.49\linewidth}
    \vspace{0pt}
    \centering
    \vspace{-13pt}
    \caption{Critical update ratio.}\label{fig:result-edge-update-50}
  \end{subfigure}
  \begin{subfigure}[t]{0.49\linewidth}
    \vspace{0pt}
    \centering
    \vspace{-13pt}
    \caption{Graph update time-series.}\label{fig:result-edge-update-merge}
  \end{subfigure}
  \vspace{-6pt}
  \caption{Update of dynamic graphs.}\label{fig:result-edge-update}
  \vspace{-15pt}
\end{figure}

\noindent\textbf{Other accelerators.}
Figure~\ref{fig:hpca-acc-sens} shows the relative performance of existing accelerators. We evaluate four designs: two for ordering (merge-sort\cite{song2016parallelhm} and insertion-sort\cite{xilinxsorting} accelerators), and two for selection (a stream-based FPGA sampler\cite{yuchen2024a} and precomputation and vector quantization\cite{yunki2025flag}).
We evaluate three configurations for each accelerator: i) \texttt{Pure}, which uses the accelerator alone and occupies 100\% of the FPGA; ii) \texttt{SCR}, which partitions the FPGA 30:70 and adds AutoGNN's SCR unit to the 30\% region; and iii) \texttt{Auto}, which subdivides the 70\% region and adds AutoGNN's UPE to one half to enable end-to-end preprocessing (akin to \texttt{AutoPre}). The final bar is \texttt{DynPre}.
\texttt{SCR}, \texttt{Auto}, and \texttt{DynPre} deliver 1.7$\times$, 3.3$\times$, and 4.5$\times$ speedups over \texttt{Pure}, respectively. \texttt{Pure} accelerates only one stage (ordering or selection), so overall performance is bounded by the remaining stages and by large transfer overheads. \texttt{SCR} speeds up most preprocessing via the added SCR, but still incurs high transfers due to repeated host-GPU-FPGA handoffs. \texttt{Auto} eliminates those transfers by performing end-to-end preprocessing on the FPGA, yet, like \texttt{AutoPre}, splits the UPE region into ordering and selection-only sub-engines, lowering LUT utilization and capping performance.


\noindent\textbf{Graph update.}
Figure~\ref{fig:result-edge-update} summarizes two views: i) the minimum graph-update ratio that perturbs GNN outputs while varying the number of layers (Figure\,\ref{fig:result-edge-update-50}), and ii) a time-series of per-hour update ratios (Figure\,\ref{fig:result-edge-update-merge}).
In \texttt{SO} and \texttt{TB}, newly added vertices have relatively low connectivity; consequently, the fraction of vertices influenced by updates stays nearly constant as layers increase. In contrast, \texttt{JR} and \texttt{AM} exhibit high connectivity for new vertices; thus, with more layers, even a small number of updates can affect most of the graph.
Motivated by these characteristics, practical services rebuild the graph once the update ratio reaches 0.5\%\cite{gao2023mega,jun2019intentgc}.
Figure~\ref{fig:result-edge-update-merge} shows the per-hour update ratios for two representative dynamic graphs, \texttt{TB} and \texttt{SO}. On average, 0.74\% of the graph changes every two hours, implying that frequent graph reconstruction is required to maintain high GNN accuracy.



\noindent\textbf{Consecutive diverse graphs.}
Many applications require real-time inference on diverse graphs that vary over time. Figure~\ref{fig:result_timeline-g} illustrates a scenario where two distinct graphs, MV and SO, arrive sequentially. \texttt{StatPre}, which utilizes only the optimal configuration of the first graph, experiences a significant throughput drop during SO preprocessing. This is because the graphs have distinct optimal configurations due to MV's large graph degree. \texttt{DynPre} reconfigures the hardware, which, although it adds a 0.23 second latency, boosts throughput by 2.9$\times$ thereafter. As a result, \texttt{DynPre} reduces the overall preprocessing time by 56\%. Figure~\ref{fig:result_mixed_timeline} further evaluates when a pair of similar/different graphs is input. Compared to \texttt{StatPre}, \texttt{DynPre} reduces the preprocessing latency of similar and different datasets by 14.6\% and 46.1\%. This highlights the hardware reconfiguration capability of \texttt{DynPre}, beneficial in scenarios where the graphs vary.



\enlargethispage{10pt}

\begin{figure}
  \vspace{-10pt}
  \begin{minipage}[t]{0.49\linewidth}
    \vspace{0pt}
      \centering
      \includegraphics[width=\linewidth]{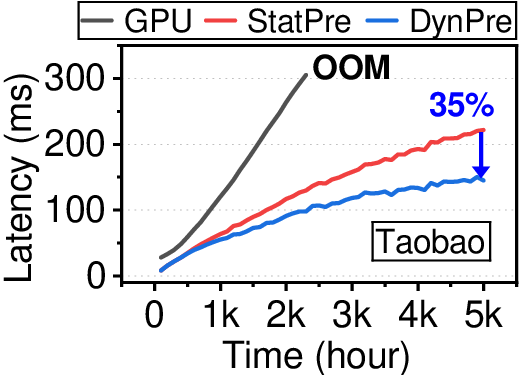}
      \vspace{-15pt}
      \caption{Dynamic graph.}\label{fig:result_dynamic_g}
  \end{minipage}%
  \begin{minipage}[t]{0.49\linewidth}
    \vspace{0pt}
      \centering
      \includegraphics[width=\linewidth]{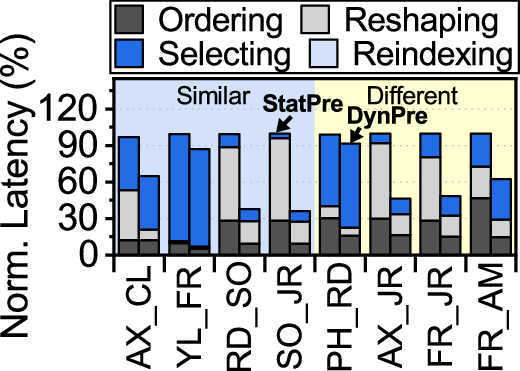}
      \vspace{-15pt}
      \caption{Mixed edges.}\label{fig:result_mixed_node}
    \end{minipage}
  \vspace{-15pt}
\end{figure}

\noindent\textbf{Dynamic graphs.}
Interactions in a social graph (SO) or item purchases in an e-commerce graph (TB) are often added over time. Figure~\ref{fig:result_dynamic_g} shows TB's end-to-end latencies, while edge count and degree increase by 112$\times$ and 9.2$\times$, respectively, over time. The performance benefits of \texttt{StatPre} over \texttt{GPU} become more pronounced, since preprocessing latency occupies a larger share over time. Compared to \texttt{StatPre}, \texttt{DynPre} further reduces the end-to-end latency by 35\% by adapting the optimal hardware configuration.


\noindent\textbf{Mixed edges.}
To evaluate concurrent inferences on two different graphs, we mix edges from graphs within the same category and across different categories. Figure~\ref{fig:result_mixed_node} shows the preprocessing latency of these mixed edges under \texttt{StatPre} and \texttt{DynPre}. Compared to \texttt{StatPre}, \texttt{DynPre} reduces preprocessing latency by 98.9\% for same-category graph mixes and by 74.1\% for cross-category graph mixes.




\section{Related Work}\label{sec:related_work}
Numerous accelerators have been proposed to reduce GNN processing latency, including GPU-based frameworks \cite{sun2022cognn, xiao2025dcgg}, FPGA-based implementations \cite{geng2020awb,zhang2021boostgcn,zhang2020hardware,chen2021rubik,zeng2020graphact,liang2020deepburning}, and domain-specific ASIC designs \cite{yan2020hygcn, chen2022regnn,kiningham2022grip,liang2020engn}. These works focus on GNN inference while ignoring the preprocessing overhead.
Although some accelerator works target sorting \cite{chen2015energy, Koch2011FPGASortAH,liu2015radixboost,saitoh2018very}, format processing \cite{sarkar2023flowgnn}, or sampling~\cite{li2022hyperscale}, they devote most resources to a single function, thus unsuitable for end-to-end GNN preprocessing.
In contrast, AutoGNN accelerates preprocessing with reconfigurable unified PEs, achieving high performance across various graphs.

\section{Conclusion}\label{sec:conclusion}
AutoGNN is an FPGA-based accelerator addressing bottlenecks in GNN preprocessing. Leveraging reconfigurable hardware and specialized components, AutoGNN accelerates tasks such as graph conversion and sampling. Our software framework ensures adaptability to varying workloads, optimizing performance in real-time. Implemented on a 7$n$m enterprise FPGA, AutoGNN delivers up to 9.0$\times$ and 2.1$\times$ speedup compared to conventional \mbox{GPU-accelerated systems}.



\section{Acknowledgements}
The authors thank anonymous reviewers of ISCA'25, MICRO'25 and HPCA'26 for their constructive feedback.
This work was supported by Samsung Research Funding \& Incubation Center for Future Technology of Samsung Electronics under Project Number SRFC-IT2302-05.
This work is protected by one or more patents. Myoungsoo Jung is the corresponding author.

\setstretch{1.0}


\bibliographystyle{IEEEtranS}
\bibliography{ref}

\end{document}